\documentclass[final,3p,times]{elsarticle}
\usepackage{float}
\usepackage{lineno,hyperref}
\usepackage{amssymb}
\usepackage{amsthm}
\usepackage{mathrsfs}
\usepackage[namelimits]{amsmath}
\usepackage{longtable}
\usepackage{threeparttable}
\usepackage{booktabs}
\hypersetup{
colorlinks=true,
linkcolor=black
}
\biboptions{authoryear}

\begin{document}
\begin{frontmatter}
\title{Integrated optimization of train timetables rescheduling and response vehicles on a disrupted metro line}

\author[1,2]{Hui Wang}
\ead{huiwang24@bjtu.edu.cn}
\author[2]{Jialin Liu}
\ead{18114007@bjtu.edu.cn}
\author[1]{Feng Li\corref{mycorrespondingauthor}}
\ead{fengli0925@bjtu.edu.cn}
\author[3]{Hao Ji}
 \ead{jihao_78@126.com}
\author[2,3]{Bin Jia \corref{mycorrespondingauthor}}
 \ead{bjia@bjtu.edu.cn}
\author[1]{Ziyou Gao}
\ead{zygao@bjtu.edu.cn}
\cortext[mycorrespondingauthor]{Corresponding author: Feng Li, Bin Jia}
%\cortext[mycorrespondingauthor]{Bin Jia}

\address[1]{State key laboratory of Rail Traffic Control and Safety, Beijing Jiaotong university, Beijing 100044, China}
\address[2]{Key laboratory of transport industry of big data application technologies of comprehensive transport, Beijing Jiaotong university, Beijing 100044, China}
\address[3]{School of Economics and Management, Xi’an Technological University, Xi’an 710021, China}
\begin{abstract}
When an unexpected metro disruption occurs, metro managers need to reschedule timetables to avoid trains going into the disruption area, and transport passengers stranded at disruption stations as quickly as possible. This paper proposes a two-stage optimization model to jointly make decisions for two tasks. In the first stage, the timetable rescheduling problem with cancellation and short-turning strategies is formulated as a mixed integer linear programming (MILP). In particular, the instantaneous parameters and variables are used to describe the accumulation of time-varying passenger flow. In the second one, a system-optimal dynamic traffic assignment (SODTA) model is employed to dynamically schedule response vehicles, which is able to capture the dynamic traffic and congestion. Numerical cases of Beijing Metro Line 9 verify the efficiency and effectiveness of our proposed model, and results show that: (1) when occurring a disruption event during peak hours, the impact on the normal timetable is greater, and passengers in the direction with fewer train services are more affected; (2) if passengers stranded at the terminal stations of disruption area are not transported in time, they will rapidly increase at a speed of more than 300 passengers per minute; (3) compared with the fixed shortest path, using the response vehicles reduces the total travel time about $7\%$. However, it results in increased travel time for some passengers.
\end{abstract}


\end{frontmatter}

%\linenumbers
\section{Introduction}
\subsection{Background}
Urban metro system plays an essential role in daily trips so that the operators need to make efficient, safe and reliable plans to meet the huge traveling demand. However, some unexpected disruptions occur now and again. Metro managers often use two management strategies to respond to such emergencies: one is to reschedule normal train timetables, and the other is to arrange shuttle vehicles to evacuate stranded passengers. For example, the line tracks between Serangoon station \& Sengkang station in the North East Line of Singapore was disrupted because of a power fault on March 27, 2021. This disruption lasted for hours, and tens of thousands of passengers were affected. To ensure the safety of trains, the Singapore Bus Service (SBS) quickly adjusted the train timetable in North East Line, and two free bus bridging routes were also opened (\cite{singapore}). However, as seen from the viewpoints of Twitter, passengers are still not satisfied with the current disruption response plan. There is still much room for improving the current timetable rescheduling and bus bridging service.

Except for the power failure, unexpected disruption might be triggered by many reasons, such as the breakdown of signals, the equipment failure of line tracks, mistakes of operators, etc (\cite{golightly2017characteristics}, \cite{pender2013disruption}).
	
\begin{figure}
		\centering 
		\includegraphics[width=\textwidth]{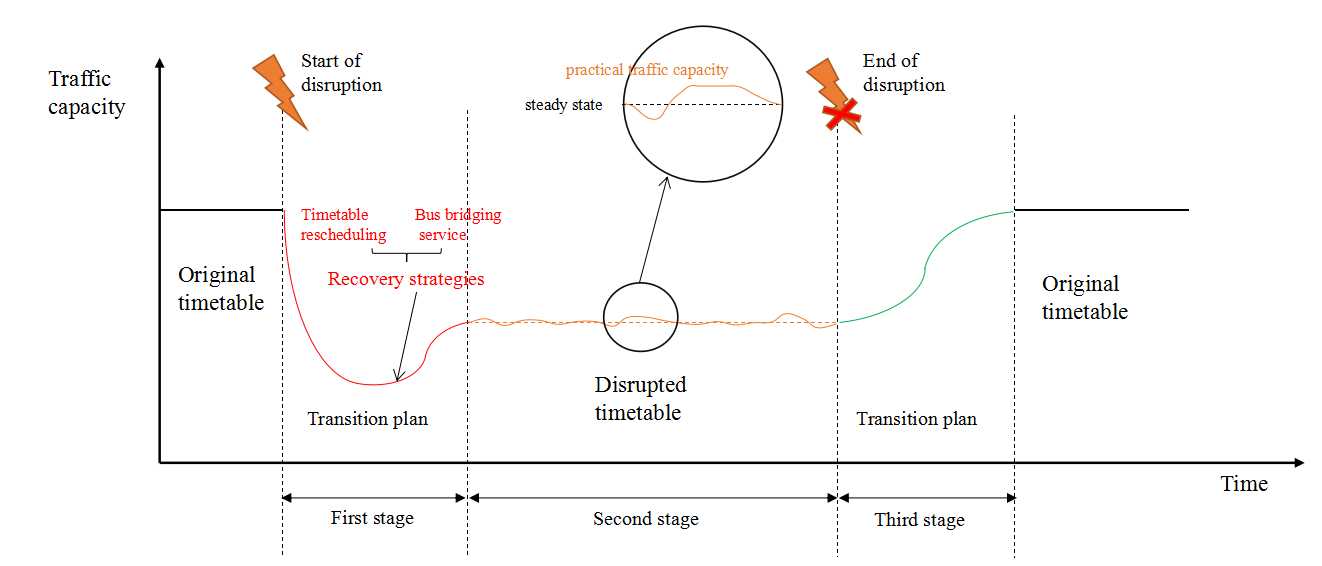}
		\caption{Traffic capacity during metro disruption period}\label{introduction1}
\end{figure}

The metro disruption period is divided into three stages (\cite{ghaemi2018macroscopic}). As shown in Figure \ref{introduction1}, in the first stage, the traffic capacity of the metro system drops dramatically, and the normal timetable is converted into a disrupted timetable. The metro operators analyze the disruption reason and predict the duration time. According to the estimated disruption duration time, the metro managers determine whether the bus bridging service should be activated, and what kind of rescheduling strategies should be executed. In the second stage, the traffic capacity maintain a lower steady state after using recovery strategies. In the third stage, the disrupted timetable is swifted to the normal timetables after disruption, and the traffic capacity would return to its normal level. To provide a complete response scheme, we focus on the first stage and aim at providing efficient recovery strategies. (\cite{cacchiani2014overview}).

Timetable rescheduling aims to avoid the trains running into the disruption area by adjusting the normal timetable. While, response vehicles aim to bridge the disrupted metro line by adding auxiliary vehicle services, which ensures that affected passengers can still reach their destinations. To the best of our knowledge, there are no studies majoring at optimizing both the metro timetable rescheduling and bus bridging service. To increase the efficiency of recover strategies, a complete plan including metro timetable rescheduling and bus bridging service are needed. Besides,the current bus bridging service can be seen as one type of response vehicles that has fixed route and departure frequency. However, it can not capture the time-varying road conditions. Thus, considering the time-varying road conditions, response vehilces are required to adjust their route choice according to the real-time road consitions.

To address the research gap, this work focuses on two questions as follows:\\
1. How to coordinate the train timetable rescheduling and response vehicle scheduling into an application? Because the transport capacity of trains is much greater than vehicles, it is a challenge task to bridge the capacity gap between both traffic modes. \\
2. How to schedule the real-time routes of response vehicles considering real-time traffic conditions?

This paper formulates a two-stage optimiztion model. In the first stage, we compute a feasible rescheduled timetable with the objective to minimize passenger waiting time, and the train safety and passenger assignment are considered. In particular, we introduce instaneous parameters and variables to calculate the passenger accumulation at each station. The result of passenger accumulation is then transferred into the demand of response vehicles. In the second stage, the real-time routes of response vehicles are guided by the SO-DTA model (\cite{Athanasios2000}). The road is discretized into a serious of successive cells, traffic signals and road capacity are both considered.
\subsection{Literature review}
\subsubsection{Literature}
Metro disruption management refers to the field of conventional railway disruption management. In the early ages, disruption management is urgent in daily railway operations, because the railway infrastructures are not complete. Potential influence factors (\cite{jespersen2009disruption}) and fault-to-incident propagation (\cite{yan2019failure}) are essential for disruption prevention in the railway lines. According to the extent of disruption, railway  disruption can be into two types: partial disruption and complete disruption (\cite{cacchiani2014overview}).
For the partially disrupted railway line, some tracks with the same direction are disrupted. Trains of different directions have to share the same track in the partially disrupted tracks. (\cite{LOUWERSE2014583}, \cite{luo2018train}). However, the railway line would be divided into two separate parts if it is completely disrupteded (\cite{doi:10.1287/trsc.2015.0618},\cite{nielsen2012rolling}). This classification is also accepted by metro disruption management. To cope with unexpected disruptions, four strategies are adopted in the metro timetable rescheduling: short-turning, skip-stopping, canceling, and speed adjustment. The short-turning strategy is to make the train circle a partial area of the railway line (\cite{ghaemi2017microscopic},\cite{schettini2022demand}). The result of \cite{ghaemi2018macroscopic} have revealed that the optimal short-turning strategy depends on both the start time and the length of the disruption. The skip-stopping strategy is to let the train not stop at some stations, as it can avoid trains loading too many passengers. As this strategy makes tiny changes to the metro timetable, it is usually used together with other strategies (\cite{ZHU2020282}). Canceling is to cancel some train services in the normal timetable (\cite{ZHU2020282}). Except for the above three strategies, the normal timetable can also be rescheduled by adjusting train speed. This strategy is from a microscopic aspect, and the train running time of railway tracks would be changed. Speed adjustment is usually adopted to reduce passenger delays (\cite{narayanaswami2013modelling}), or to achieve real-time running control of high-speed railway trains (\cite{xu2017train}). 

Most literature focus on operator-oriented timetable rescheduling. The objectives of these literature usually includes cost of rescheduling strategies (\cite{schettini2022demand}), train delay (\cite{tessitore2022simulation}, \cite{xu2016rescheduling}), or energy consumption (\cite{bueno2020analytical}). Few papers focus on integrating timetable rescheduling with other topics.i.g.,\cite{wang2021real} integrates rolling stock circulation and timetable rescheduling. \cite{bevsinovic2021matheuristic} incorporates train timetable rescheduling and passenger flow control. These literature do not conclude the interests of passengers, which is not enough for providing high-quality services. To compensate for this shortcoming, some recent literature focus on passenger-oriented timetable rescheduling. As the disruption information is unknown, passengers have to decide whether to wait until the disruption has ended (\cite{2014Waiting}, \cite{ZHANG2018409}) or board other transport means (\cite{2018The}, \cite{2018Impact}, \cite{xu2021optimizing}). The choice of passengers depends on their general cost, their available incomes (\cite{pnevmatikou2015metro}), and so on. To better fit the demand of passengers, passenger-oriented timetable rescheduling is now accepted by scholars. The objective of passenger-oriented timetable rescheduling is to minimize the passenger total traveling time (\cite{ZHU2020282}, \cite{cite-key3}), or the satisfaction of passengers (\cite{BINDER201778},\cite{huang2020coupling}). Usually, these problems are solved by decomposition algorithms (\cite{zhan2021integrated}, \cite{gao2016rescheduling}) or intelligent algorithms (\cite{zilko2016modeling}). However, these literature does not consider the substitute transport, and passengers have to wait during thr overall disruption. 

Owing to the blockage of disrupted metro line, some passengers can not successfully reach their destination by metro. Thus, auxiliary transport is needed to serve these passengers. In practical operations, buses have been adopted as their advantage of large capacity (\cite{singapore}). For example, the first train timetabling (\cite{guo2019first}. \cite{kang2021first}), last train timetabling (\cite{2017Two}, \cite{yang2020collaborative}), and contract design (\cite{zhang2020metro}). During peak hours, the bus bridging service can also be applied to assist passenger control on a busy metro line (\cite{https://doi.org/10.1111/mice.12265}). While facing with a long-time disruption, the operator has to desgin the bus bridging routes and their frequencies (\cite{Kepaptsoglou2009}, \cite{doi:10.1287/trsc.2014.0577}). For a disrupted single metro line, most literature focus on the design of bus bridging to minimize the cost of passengers (\cite{wang2019optimization}, \cite{doi:10.1287/trsc.2015.0647}). While, for a disrupted metro network, the attention are more paid to the resilience to potential disruptions (\cite{2014Enhancing}), or to build a integrated rail-bus transit network (\cite{hua2018effect}, \cite{tan2020evacuating}, \cite{dou2019parallel}). 

However, there are two shortcomings of the current bus bridging research. First, most literature assumes that the auxiliary bus has fixed routes and fixed frequencies. Some researchers have also considered breaking the mindset of bus bridging services, and they have designed a more flexible bus bridging plan in response to metro disruptions (\cite{wang2016feeder}, \cite{GU2018209}). Second, current research on the bus bridging service assumes that the road condition is constant all the time, and the bus can reach each station on time. In daily operation, this assumption is a rather strong hypothesis, as it has not considered the variation of road traffic. For example, bus bunching is a common phenomenon in our daily life, which will lead to an unstable bus delay on the congested road (\cite{daganzo2009headway}, \cite{daganzo2011reducing}, \cite{delgado2012much}, \cite{petit2018dynamic}). Therefore, to better fit the practical operational demand, the real-time road traffic information should be concluded in our reponse schemes.

To capture the real traffic dynamics, the DTA technologies (\cite{peeta2001foundations}, \cite{szeto2006dynamic}) are considered in our research. Usually, it can be classified into two types: simulation-based approaches (\cite{sbayti2007efficient},\cite{levin2015improving}, \cite{shafiei2018calibration}) and analytical approaches (\cite{2012Dynamic}, \cite{2013A}). The simulation-based models often suffer from a huge computational burden, and the analytical approach also has such a deficiency because the models are often written as path-based formulations. Moreover, the analytical approach has the problem of poor convergence in large-scale networks (\cite{2016Multi}).
Fortunately, a linear programming (LP) framework for system optimal DTA (SO-DTA) problem (\cite{Athanasios2000}) was proposed, in which the cell transmission model (CTM) (\cite{DAGANZO1994269}, \cite{DAGANZO199579}) is employed to load traffic flow. This framework is widely used in large-scale emergency traffic planning and organization due to its lower computational burden. (\cite{2007Modeling}, \cite{2009Evacuation}, \cite{XIE2010295}, \cite{BENTAL20111177}, \cite{KIMMS20181122}, \cite{2014Dynamic}, \cite{liu2021lane} ). These literature has provided the foundation for the design of response vehicles on the road.
 
\subsubsection{Our focus}
As mentioned above, most literature focuses on either train timetable rescheduling or bus bridging service. As far as the author knows, there is no literature integrating the metro timetable rescheduling and bus bridging service. Just like \cite{ZHU2020282}, the metro timetabling rescheduling does not take the substitute transport take into account, they assume that the passengers can wait until the disruption has ended. While in the design of the bus bridging service, the metro timetable is assumed to be fixed (see \cite{doi:10.1287/trsc.2014.0577}, etc.). Thus, in this paper, we aim to integrate both the metro timetable rescheduling and response vehicle scheduling.  

Besides, the modeling method of timetable rescheduling can be extended to deal with different problems. \cite{ZHU2020282} has proposed an improved Dynamic Altenative Graph (DAG) formulation to reschedule the train timetable, and the passenger assignment procedure is to assign passengers to the alternative graph. \cite{huang2020coupling} use big-M and time-indexed formulations to reschedule the metro timetable, and the pasenger accumulation is measured by trains. However, these two important literature can not capture the time-varying accumulation of passengers, and this is required in design of response vehicles. To observe the real-time passenger accumulation in a station, we have introduced instaneous parameters and variables to record the passenger trajectory. Although there are four general reshceduling strategies, the short-turning and canceling strategy are efficient in response to unexpected disruption (\cite{ZHU2020282}). To simplify the operation of metro operators, our paper only includes these two main strategies. 
 	\begin{figure}[H]
		\centering 
		\includegraphics[width=0.8\textwidth]{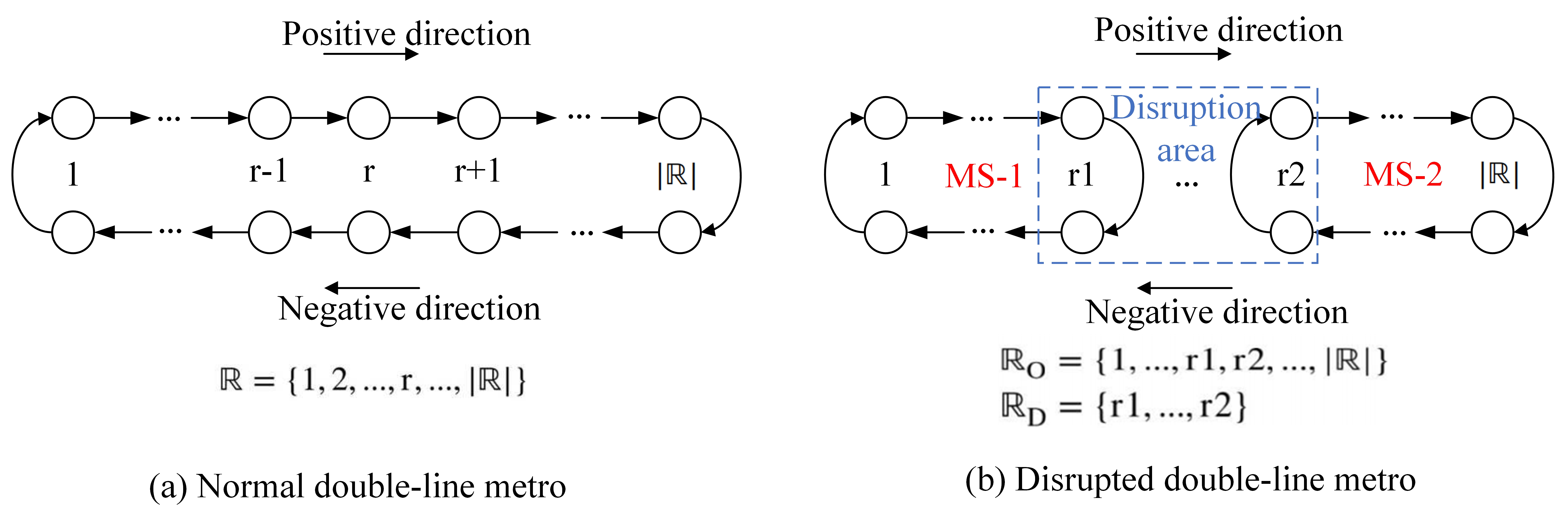}
		\caption{Double-line metro under disruption scenario}\label{metro_line}
	\end{figure}
In the field of metro disruption management, DTA is a new technique in response to the disruption. Compared with traditional optimization method (i.g., \cite{Kepaptsoglou2009}, \cite{doi:10.1287/trsc.2014.0577}), the biggest advantage of DTA is that it can capture real-time information of vehicles on the road. 

Therefore, this paper contributes to the literature through three aspects:

$\bullet$A new two-stage optimization model is built to coordinate the metro timetable rescheduling and response vehicle scheduling. In particular, the passenger accumulation is measured by our proposed model, which provides a new approach to passenger-oriented timetabling compared with current research.

$\bullet$Most literature take bus bridging service as a substitute for trains, we explore to provide more flexible service by scheduling response vehicles. This provides a new perspective to the response of metro disruption.

$\bullet$The application of SO-DTA can capture the real-time road traffic, which is closer to reality compared with current metro disruption research.

The remainder of this paper is organized as follows. In section 2, the problem statements of our research are described. Section 3 presents a MILP model to reschedule the metro timetable. Furthermore, the real-time routes of response vehicles are optimized based on the SO-DTA model in section 4. Numerical cases of Beijing Metro Line 9 are conducted in section 5. Finally, the conclusions and possible topics for future research have been presented in section 6. 
 	\begin{figure}[H]
		\centering 
		\includegraphics[width=\textwidth]{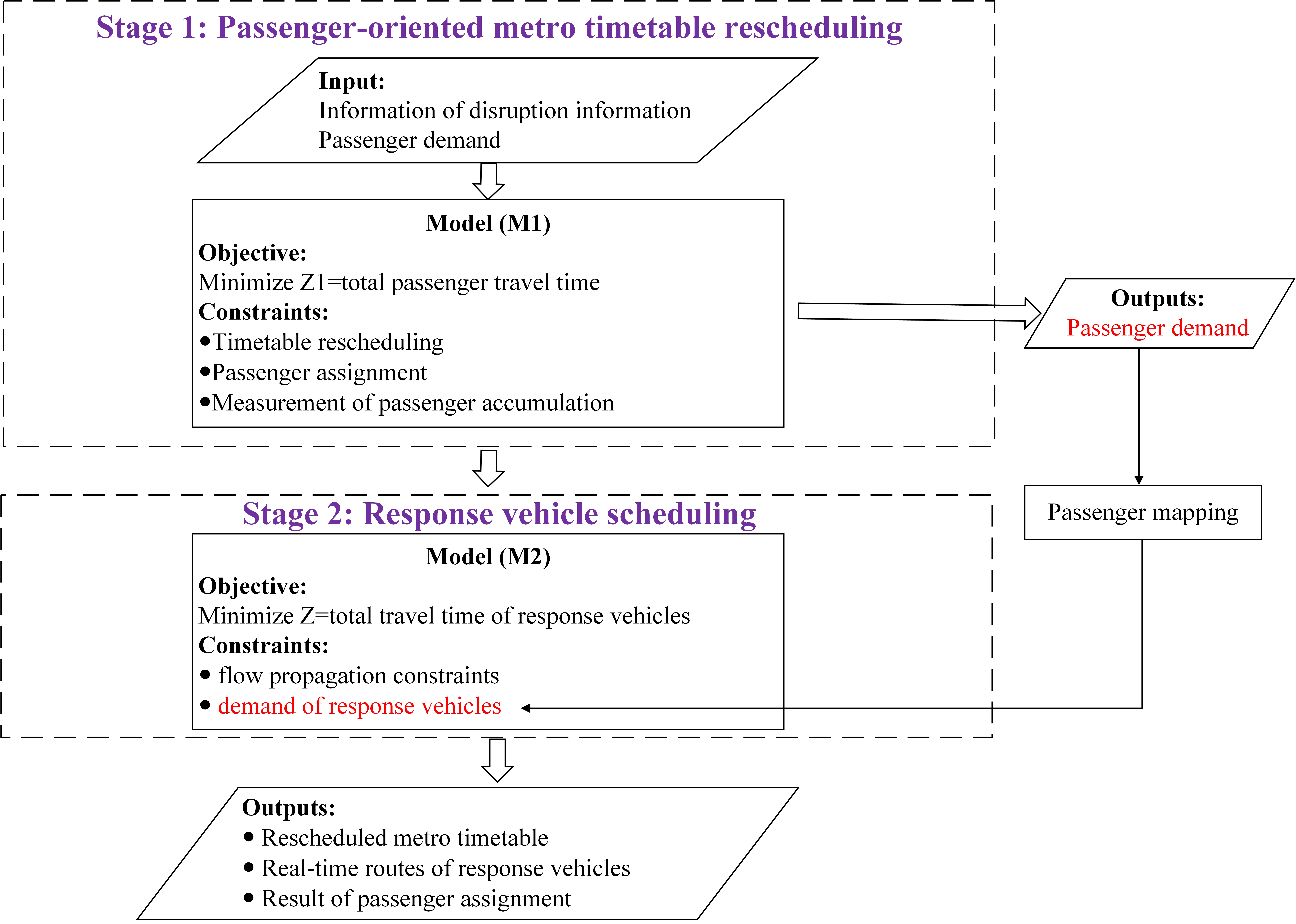}
		\caption{Overview of two-stage optimization model}\label{model_overview}
	\end{figure}
\section{Problem description}
We investigate the problem of metro timetable rescheduling and response vehicle scheduling on a disrupted double-line metro in this research. A double-line metro is made up of two separate lines that run in opposite directions. Figure \ref{metro_line} (a) shows that we use the set $\mathbb{R}$ to denote all of the stations in the normal metro line, and the line track is described as a tuple (r, r+1) or (r, r-1). Note that the tuple (r, r+1) denotes the line track in the positive direction, while (r, r-1) denotes the line track in the negative direction. Trains circulate between the normal metro line's terminal stations (Station 1 and Station $|\mathbb{R}|$). However, due to an unexpected disruption, the metro line would be divided into two parts. These two parts are called as MS-1 and MS-2, and they are referred to as \textbf{operational areas} because trains can operate in them. As shown in Figure \ref{metro_line}, an unexpected disruption occurred between stations r1 and r2. The operational area is denoted by $\mathbb{R}_{\rm{O}}=\{1,...,\rm{r1},\rm{r2}..., |\mathbb{R}|\}$. MS-1 denotes the stations from Station 1 to r1, while MS-2 denotes the stations from Station 2 to $|\mathbb{R}|$. Furthermore, we refer to the area between Stations r1 and r2 as the \textbf{disruption area} because trains cannot pass through it during a disruption scenario. As a result, trains will be rescheduled in the operational area while response vehicles will be scheduled in the disruption area in this paper.

In comparison to previous studies on metro disruption, we incorporate metro timetable rescheduling and response vehicle scheduling. The modeling methodologies for metro timetable rescheduling and response vehicle scheduling, however, differ. For the metro timetable rescheduling, train services in the normal timetable are rescheduled to avoid conflicts with spatio-temporal disruption area (definitions see as \ref{definitions}). Passengers are assigned to available trains in the operational area; For the response vehicle scheduling, real-time routes of response vehicles are modeled using the method of dynamic traffic assignment, and vehicles are loaded into the road network in the disruption area. How can we bridge the gap between metro trains and response vehicles given the topological differences between metro lines and road networks?

To include both the metro timetable rescheduling and response vehicle scheduling in a unified optimization structure, we built a two-stage optimization model that first reschedules the normal metro timetable and then optimizes the real-time routes of response vehicles. An overview of the two-stage optimization model is shown in Figure \ref{model_overview}. 

Given the disruption information on a railway line and the passenger demand, we obtain a passenger-oriented rescheduled timetable in the first stage. In the first stage, the metro timetable rescheduling problem is described as a MILP, with the goal of minimizing total passenger waiting time. We consider three constraint modules in this stage: 1) timetable rescheduling. The timetable rescheduling aims to reschedule the normal timetable by using two strategies: cancellation and turnaround. Besides, the train's safety in the new timetable is also ensured by the headway constraints. 2) passenger assignment. Passengers are restricted to be assigned to available train services. 3) measurement of passenger accumulation. The instantaneous parameters and variables are introduced to describe the time-varying passenger accumulation; In the second stage, the response vehicle scheduling problem is described as a LP, with the goal of minimizing the systematic total travel time of response vehicles. Before the scheduling of response vehicles, the real-time passenger demands are transferred into the demands of response vehicles by the procedure of passenger mapping. The constraints are referred to \cite{Athanasios2000}. We finally obtain an optimally rescheduled metro timetable, real-time routes of response vehicles, and passengers are transported to their destinations or transfer stations that can reach their destination through the two steps of optimizing models M1 and M2.

In order to ensure the rigorism of our focused problem, some necessary assumptions should be declared before the proposition of our model:

(1) The duration of disruption can be predicted. Much of the metro disruption literature accepts this assumption (\cite{wang2021real},\cite{ZHANG2018409},\cite{huang2020coupling}). Once the disruption has occurred, the metro operators should analyze the cause of the disruption and estimate the duration of the disruption. Response strategies are based on the estimated disruption period for practical operations.

(2) The metro stations (except for the terminal stations in the disruption area) in the disruption area are closed during the disruption period. This assumption refers to the practical disruption case of Singapore on March 27, 2021. As the trains can not go through the disruption area, and the passengers can not board trains in this area. Thus, for the sake of recovery work on metro line tracks, these stations are closed. 

(3) The terminal stations in the disruption area can provide trains with turnaround operations. To ensure the safe driving of trains, the train would turn around at the nearest turn-back station from disrupted terminal stations. Although some tracks are not broken down, they are also regarded as disrupted tracks because trains can not run across them.

(4) The response vehicles are sufficient. As the response vehicles can be deployed from nearby car or bus depots, we assume that sufficient vehicles can be deployed swiftly from these depots. This assumption has ensured that all the passenger demand in the disruption area can be boarded by the response vehicles.

(5) The response vehicles have the same holding capacity. This assumption was made to simplify the mathematical model, and the response vehicles are divided into several fleets.

(6) Special lanes are planned for the response vehicles. For each road, there are two special lanes with different directions planned for the response vehicles.

\section{Stage 1: timetable rescheduling}
In this section, we develop a MILP model for passenger-oriented timetabling with the objective of minimizing total passenger waiting time. The MILP model is made up of three constraint modules: 1) timetable rescheduling, 2) passenger assignment, and 3) passenger accumulation measurement. The following is our mathematical model:
\subsection{Objective function}
The objective is to minimize the total waiting time of passengers, as well as to penalize the unassigned passengers, which is

\begin{equation}
\min Z_{1}=\sum_{{\rm{p}\in  \mathbb{P}}}\sum\limits_{{\rm{u}}\in \mathbb{U}\cup\mathbb{\tilde{U}}} \left( \begin{matrix}\breve{{\rm{w}}}_{{\rm{p}}}^{{\rm{u}}}x_{{\rm{p}}}^{{\rm{u}}}+{\rm{M}}x_{{\rm{p}}}\end{matrix} \right) \qquad\qquad
\end{equation}
where the parameter $\breve{{\rm{w}}}_{{\rm{p}}}^{{\rm{u}}}$ denotes the waiting time if passenger flow p takes the train service u.i,.e., $\breve{{\rm{w}}}_{{\rm{p}}}^{{\rm{u}}}={\rm{\hat{t}}}_{{\rm{u,{\rm{\hat{o}_{p}}}}}}^{a}-{\rm{\hat{t}_{p}}}$. The first term of objective function measures the total waiting time of passenger flows, which ensures the total waiting time of passengers can be minimized. While the last term penalizes the failure boarding of passengers, which ensures the passengers are assigned to trains as many as possible.\\
\textbf{Remark 1.} The big-M parameter in the objective function $Z_{1}$ should satisfy: M $\ge \max\limits_{{\rm{p}\in \mathbb{P}}, {\rm{u}}\in \mathbb{U}\cup\mathbb{\tilde{U}}}\{\breve{{\rm{w}}}_{{\rm{p}}}^{{\rm{u}}}\}$; otherwise, the passenger might not be assigned completely.\\
\textbf{Proof.} see as \ref{proof1}.
\subsection{Constraints}
In the following, the constraints for train service activation, station identification, train headway, passenger assignment, train capacity, an passenger accumulation are introduced successively.
\subsubsection{Activation of normal train services}
The metro timetable includes the train route as well as station arrival and departure times. The normal timetable is generated using daily passenger demand data. However, if an unexpected disruption occurs on the metro line, the normal timetable should be adjusted to ensure train safety. That is, in the disruption scenario, the train should not pass through the spatio-temporal disruption area and should turn around at the disruption area's terminal station. If the train service in the normal timetable conflicts with the spatio-temporal disruption area, the train trajectory should be revised. The following constraints depict the activation of normal train services:

\begin{subequations}
\begin{equation}
s.t.\left( \begin{matrix}t_{{\rm{u,r}}}^{a}\\ t_{{\rm{u,r}}}^{d}\end{matrix} \right) \ge \left( \begin{matrix}{\rm{t}}_{{\rm{u,r}}}^{a}\\ {\rm{t}}_{{\rm{u,r}}}^{d}\end{matrix} \right)-{\rm{M}} (1-a_{{\rm{u}}}+\Theta_{{\rm{u}}}), \forall  {\rm{u}\in  \mathbb{U}}\cup\mathbb{\tilde{U}}, {\rm{r}\in  \mathbb{R}},\label{normal1}\qquad \qquad 
\end{equation}
\begin{equation}
\quad \ \left( \begin{matrix}t_{{\rm{u,r}}}^{a}\\ t_{{\rm{u,r}}}^{d}\end{matrix} \right) \le \left( \begin{matrix}{\rm{t}}_{{\rm{u,r}}}^{a}\\ {\rm{t}}_{{\rm{u,r}}}^{d}\end{matrix} \right)+{\rm{M}} (1-a_{{\rm{u}}}+\Theta_{{\rm{u}}}), \forall  {\rm{u}\in  \mathbb{U}}\cup\mathbb{\tilde{U}}, {\rm{r}\in  \mathbb{R}},\label{normal2}\qquad \qquad 
\end{equation}
\end{subequations}
\begin{subequations}
\begin{equation}
\quad \qquad \quad \left( \begin{matrix}t_{{\rm{u,r}}}^{a}\\ t_{{\rm{u,r}}}^{d}\end{matrix} \right) \ge \left( \begin{matrix}{\rm{t}}_{{\rm{u,r}}}^{a}\\ {\rm{t}}_{{\rm{u,r}}}^{d}\end{matrix} \right)+{\rm{M}} (a_{{\rm{u}}}+\Theta_{{\rm{u}}}+{\rm{f}}_{{\rm{u}}}-3), \forall  {\rm{u}\in  \mathbb{U}}, {\rm{r}\in  \mathbb{R}_{O}^{1}}\setminus  \mathbb{R}_{\rm{T}}^{1},\label{timetable2_1}\qquad \qquad 
\end{equation}
\begin{equation}
\qquad \qquad\ \left( \begin{matrix}t_{{\rm{u,r}}}^{a}\\ t_{{\rm{u,r}}}^{d}\end{matrix} \right) \le \left( \begin{matrix}{\rm{t}}_{{\rm{u,r}}}^{a}\\ {\rm{t}}_{{\rm{u,r}}}^{d}\end{matrix} \right)+{\rm{M}} (3-a_{{\rm{u}}}+\Theta_{{\rm{u}}}-{\rm{f}}_{{\rm{u}}}), \forall  {\rm{u}\in  \mathbb{U}}, {\rm{r}\in  \mathbb{R}_{O}^{1}}\setminus  \mathbb{R}_{\rm{T}}^{1},\label{timetable2_2}\qquad \qquad 
\end{equation}
\end{subequations}
\begin{subequations}
\begin{equation}
\qquad \ \left( \begin{matrix}t_{{\rm{u,r}}}^{a}\\ t_{{\rm{u,r}}}^{d}\end{matrix} \right) \ge \left( \begin{matrix}{\rm{t}}_{{\rm{u,r}}}^{a}\\-1\end{matrix} \right)+{\rm{M}} (\Theta_{{\rm{u}}}+{\rm{f}}_{{\rm{u}}}+a_{{\rm{u}}}-3), \forall  {\rm{u}\in  \mathbb{U}}, {\rm{r}\in \mathbb{R}_{\rm{T}}^{1}},\label{timetable3_1}\qquad \qquad 
\end{equation}
\begin{equation}
\qquad \ \left( \begin{matrix}t_{{\rm{u,r}}}^{a}\\ t_{{\rm{u,r}}}^{d}\end{matrix} \right) \le \left( \begin{matrix}{\rm{t}}_{{\rm{u,r}}}^{a}\\ -1\end{matrix} \right)+{\rm{M}} (3-\Theta_{{\rm{u}}}-{\rm{f}}_{{\rm{u}}}-a_{{\rm{u}}}), \forall  {\rm{u}\in  \mathbb{U}}, {\rm{r}\in \mathbb{R}_{\rm{T}}^{1}},\label{timetable3_2}\qquad \qquad 
\end{equation}
\end{subequations}
\begin{subequations}
\begin{equation}
\qquad \quad \left( \begin{matrix}t_{{\rm{u,r}}}^{a}\\ t_{{\rm{u,r}}}^{d}\end{matrix} \right) \ge -1+{\rm{M}} (a_{{\rm{u}}}+\Theta_{{\rm{u}}}+{\rm{f}}_{{\rm{u}}}-3), \forall  {\rm{u}\in  \mathbb{U}}, {\rm{r}\in \mathbb{R}\setminus \mathbb{R}_{\rm{O}}^{1}},\label{timetable4_1}\qquad \qquad 
\end{equation}
\begin{equation}
\qquad \quad\left( \begin{matrix}t_{{\rm{u,r}}}^{a}\\ t_{{\rm{u,r}}}^{d}\end{matrix} \right) \le -1+{\rm{M}} (3-a_{{\rm{u}}}-\Theta_{{\rm{u}}}-{\rm{f}}_{{\rm{u}}}), \forall  {\rm{u}\in  \mathbb{U}}, {\rm{r}\in \mathbb{R}\setminus \mathbb{R}_{\rm{O}}^{1}},\label{timetable4_2}\qquad \qquad 
\end{equation}
\end{subequations}
\begin{subequations}
\begin{equation}
\qquad \qquad \left( \begin{matrix}t_{{\rm{u,r}}}^{a}\\ t_{{\rm{u,r}}}^{d}\end{matrix} \right) \ge \left( \begin{matrix}{\rm{t}}_{{\rm{u,r}}}^{a}\\ {\rm{t}}_{{\rm{u,r}}}^{d}\end{matrix} \right)-{\rm{M}} (2-\Theta_{{\rm{u}}}-a_{{\rm{u}}}+{\rm{f}}_{{\rm{u}}}), \forall  {\rm{u}\in  \mathbb{U}}, {\rm{r}\in  \mathbb{R}_{O}^{2}}\setminus  \mathbb{R}_{\rm{T}}^{2},\label{timetable5_1}\qquad \qquad 
\end{equation}
\begin{equation}
\qquad \qquad \left( \begin{matrix}t_{{\rm{u,r}}}^{a}\\ t_{{\rm{u,r}}}^{d}\end{matrix} \right) \le \left( \begin{matrix}{\rm{t}}_{{\rm{u,r}}}^{a}\\ {\rm{t}}_{{\rm{u,r}}}^{d}\end{matrix} \right)+{\rm{M}} (2-\Theta_{{\rm{u}}}-a_{{\rm{u}}}+{\rm{f}}_{{\rm{u}}}), \forall  {\rm{u}\in  \mathbb{U}}, {\rm{r}\in  \mathbb{R}_{O}^{2}}\setminus  \mathbb{R}_{\rm{T}}^{2},\label{timetable5_2}\qquad \qquad 
\end{equation}
\end{subequations}
\begin{subequations}
\begin{equation}
\quad \ \left( \begin{matrix}t_{{\rm{u,r}}}^{a}\\ t_{{\rm{u,r}}}^{d}\end{matrix} \right) \ge \left( \begin{matrix}{\rm{t}}_{{\rm{u,r}}}^{a}\\-1\end{matrix} \right)-{\rm{M}} (2-a_{{\rm{u}}}-\Theta_{{\rm{u}}}+{\rm{f}}_{{\rm{u}}}), \forall  {\rm{u}\in  \mathbb{U}}, {\rm{r}\in \mathbb{R}_{\rm{T}}^{2}},\label{timetable6_1}\qquad \quad 
\end{equation}
\begin{equation}
\quad \ \left( \begin{matrix}t_{{\rm{u,r}}}^{a}\\ t_{{\rm{u,r}}}^{d}\end{matrix} \right) \le \left( \begin{matrix}{\rm{t}}_{{\rm{u,r}}}^{a}\\ -1\end{matrix} \right)+{\rm{M}} (2-a_{{\rm{u}}}-\Theta_{{\rm{u}}}+{\rm{f}}_{{\rm{u}}}), \forall  {\rm{u}\in  \mathbb{U}}, {\rm{r}\in \mathbb{R}_{\rm{T}}^{2}},\label{timetable6_2}\qquad \quad 
\end{equation}
\end{subequations}
\begin{subequations}
\begin{equation}
\quad \ \left( \begin{matrix}t_{{\rm{u,r}}}^{a}\\ t_{{\rm{u,r}}}^{d}\end{matrix} \right) \ge -1-{\rm{M}} (2-a_{{\rm{u}}}-\Theta_{{\rm{u}}}+{\rm{f}}_{{\rm{u}}}), \forall  {\rm{u}\in  \mathbb{U}}, {\rm{r}\in \mathbb{R}\setminus \mathbb{R}_{\rm{O}}^{2}},\label{timetable7_1}\qquad 
\end{equation}
\begin{equation}
\quad \ \left( \begin{matrix}t_{{\rm{u,r}}}^{a}\\ t_{{\rm{u,r}}}^{d}\end{matrix} \right) \le -1+{\rm{M}} (2-a_{{\rm{u}}}-\Theta_{{\rm{u}}}+{\rm{f}}_{{\rm{u}}}), \forall  {\rm{u}\in  \mathbb{U}}, {\rm{r}\in \mathbb{R}\setminus \mathbb{R}_{\rm{O}}^{2}},\label{timetable7_2}\qquad 
\end{equation}
\end{subequations}
\begin{subequations}
\begin{equation}
 \left( \begin{matrix}t_{{\rm{u,r}}}^{a}\\ t_{{\rm{u,r}}}^{d}\end{matrix} \right) \ge -1-{\rm{M}}a_{{\rm{u}}}, \forall  {\rm{u}\in \mathbb{U}}, {\rm{r}\in  \mathbb{R}},\label{timetable8_1}\qquad \qquad \qquad \qquad \quad
\end{equation}
\begin{equation}
\left( \begin{matrix}t_{{\rm{u,r}}}^{a}\\ t_{{\rm{u,r}}}^{d}\end{matrix} \right) \le -1+{\rm{M}}a_{{\rm{u}}}, \forall  {\rm{u}\in \mathbb{U}}, {\rm{r}\in  \mathbb{R}},\label{timetable8_2}\qquad \qquad \qquad \qquad \quad
\end{equation}
\end{subequations}
\begin{equation}
 a_{{\rm{u}}}\ge1-\Theta_{{\rm{u}}},\forall {\rm{u}\in  \mathbb{U}},\label{timetable9} \qquad \qquad\qquad \qquad \qquad\qquad \qquad
\end{equation}
where binary parameter $\Theta_{{\rm{u}}}$ with value 1 indicating that train service u $\in \mathbb{U}$ conflicts with the spatio-temporal disruption area, and 0 otherwise. Constraint (\ref{normal1})-(\ref{normal2}) means that the rescheduled arrival/departure times of a train service remain unchanged if it is activated, and does not conflict with the spatio-temporal disruption area. Constraint (\ref{timetable2_1})-(\ref{timetable2_2}) mean that the arrival/departure time of an activated positive train service u $\in \mathbb{U}$ at station ${\rm{r}\in \mathbb{R}_{O}^{1}}\setminus \mathbb{R}_{\rm{T}}^{1}$ remains unchangesd if it conflicts with spatio-temporal disruption area. 
Constraint (\ref{timetable3_1}) means that the arrival times of an actived positive train service u $\in \mathbb{U}$ at terminal station ${\rm{r}\in \mathbb{R}_{\rm{T}}^{1}}$ remain unchanged if it conflicts with the spatio-temporal disruption area.
 Constraint (\ref{timetable3_2})  means that the departure time of an activated positive train service u $\in \mathbb{U}$ at terminal station ${\rm{r}\in \mathbb{R}_{T}^{1}}$ equals -1 if it conflicts with the spatio-temporal disruption area.  
Constraint (\ref{timetable4_1})-(\ref{timetable4_2}) mean that the arrival/departure time of an activated positive train service u $\in \mathbb{U}$ at station ${\rm{r}\in \mathbb{R}\setminus \mathbb{R}_{\rm{O}}^{1}}$ equals -1 if it conflicts with the spatio-temporal disruption area.

For an activated negative train service, constraints (\ref{timetable5_1})-(\ref{timetable5_2}) mean that the arrival/departure time of an activated train service u $\in \mathbb{U}$ at station ${\rm{r}\in \mathbb{R}_{O}^{2}}\setminus \mathbb{R}_{\rm{T}}^{2}$ remains unchanged if it conflicts with the spatio-temporal disruption area. 
Constraint (\ref{timetable6_1}) means that the arrival time of an actived positive train service u $\in \mathbb{U}$ at terminal station ${\rm{r}\in \mathbb{R}_{\rm{T}}^{2}}$ remains unchanged if it conflicts with the spatio-temporal disruption area.
Constraint (\ref{timetable6_2}) mean that the departure time of an activated positive train service u $\in \mathbb{U}$ at terminal station ${\rm{r}\in \mathbb{R}_{T}^{2}}$ equals -1 if it conflicts with the spatio-temporal disruption area.   
Constraints (\ref{timetable7_1})-(\ref{timetable7_2}) mean that the arrival/departure time of an activated negative train service u $\in \mathbb{U}$ at station ${\rm{r}\in \mathbb{R}\setminus \mathbb{R}_{\rm{O}}^{2}}$ equals -1 if it conflicts with the spatio-temporal disruption area. Note that, the arrival/departure time of a train service equals -1 if it is not activated (constraints (\ref{timetable8_1})-(\ref{timetable8_2})). Constraint (\ref{timetable9}) means that train service u $\in \mathbb{U}$ should be activated if it does not conflict with the spatio-temporal disruption area. 
\subsubsection{Activation of turnaround train services}
Recall that turnaround train services are saved in the set $\mathbb{\tilde{U}}$ and are activated by an activated normal train service if they conflict with the spatio-temporal disruption area. The following constraints depict the activation of turnaround train services:

\begin{subequations}
\begin{equation}
\quad \ a_{{\rm{v}}}\ge 1+(a_{{\rm{u}}}+\Theta_{{\rm{u}}}+{\rm{f}}_{{\rm{u}}}+\delta_{{\rm{u,v}}}^{{\rm{r}}}-4),\forall  {\rm{u}\in  \mathbb{U}}, {\rm{v}\in \mathbb{\tilde{U}}},{\rm{r}\in  \mathbb{R}_{T}^{1}},\label{turnaround1}
\end{equation}
\begin{equation}
\quad \ a_{{\rm{v}}}\ge 1+(a_{{\rm{u}}}+\Theta_{{\rm{u}}}-{\rm{f}}_{{\rm{u}}}+\delta_{{\rm{u,v}}}^{{\rm{r}}}-3),\forall  {\rm{u}\in  \mathbb{U}}, {\rm{v}\in \mathbb{\tilde{U}}},{\rm{r}\in  \mathbb{R}_{T}^{2}},\label{turnaround2}
\end{equation}
\end{subequations}
\begin{subequations}
\begin{equation}
\quad \ \left( \begin{matrix}t_{{\rm{u,r}}}^{a}\\ t_{{\rm{u,r}}}^{d}\end{matrix} \right) \ge \left( \begin{matrix}{\rm{t}}_{{\rm{u,r}}}^{a}\\ {\rm{t}}_{{\rm{u,r}}}^{d}\end{matrix} \right)-{\rm{M}} (1-a_{{\rm{u}}}), \forall  {\rm{u}\in \mathbb{\tilde{U}}}, {\rm{r}\in  \mathbb{R}},\label{turnaround3} \qquad \qquad\qquad\quad
\end{equation}
\begin{equation}
\quad \ \left( \begin{matrix}t_{{\rm{u,r}}}^{a}\\ t_{{\rm{u,r}}}^{d}\end{matrix} \right) \le \left( \begin{matrix}{\rm{t}}_{{\rm{u,r}}}^{a}\\ {\rm{t}}_{{\rm{u,r}}}^{d}\end{matrix} \right)+{\rm{M}} (1-a_{{\rm{u}}}), \forall  {\rm{u}\in \mathbb{\tilde{U}}}, {\rm{r}\in  \mathbb{R}},\label{turnaround4} \qquad \qquad\qquad \quad
\end{equation}
\end{subequations}
\begin{subequations}
\begin{equation}
\quad \ \left( \begin{matrix}t_{{\rm{u,r}}}^{a}\\ t_{{\rm{u,r}}}^{d}\end{matrix} \right) \ge -1-{\rm{M}}a_{{\rm{u}}}, \forall  {\rm{u}\in \mathbb{\tilde{U}}}, {\rm{r}\in  \mathbb{R}},\label{turnaround5}\qquad \qquad\qquad\qquad\qquad
\end{equation}
\begin{equation}
\quad \ \left( \begin{matrix}t_{{\rm{u,r}}}^{a}\\ t_{{\rm{u,r}}}^{d}\end{matrix} \right) \le -1+{\rm{M}}a_{{\rm{u}}}, \forall  {\rm{u}\in \mathbb{\tilde{U}}}, {\rm{r}\in  \mathbb{R}},\label{turnaround6}\qquad \qquad\qquad\qquad\qquad
\end{equation}
\end{subequations}
\begin{equation}
\quad \ \sum_{{\rm{u1}\in \mathbb{U}}}\sum_{{\rm{r}\in  \mathbb{R}}}\delta_{{\rm{u1,v}}}^{{\rm{r}}}a_{{\rm{v}}}\le \left( \begin{matrix}a_{{\rm{u}}}\\ \Theta_{{\rm{u}}}\end{matrix} \right),\forall  {\rm{u}\in \mathbb{U}},{\rm{v}\in \mathbb{\tilde{U}}}, {\rm{r}\in  \mathbb{R}},\label{turnaround7}\qquad\qquad\quad
\end{equation}
where binary parameter $\delta_{{\rm{u,v}}}^{{\rm{r}}}$ with value 1 indicating that train service ${\rm{v}\in \mathbb{\tilde{U}}}$ is activated by the turnaround operations of train service $ {\rm{u}\in  \mathbb{U}}$, and 0 otherwise. Constraints (\ref{turnaround1})-(\ref{turnaround2}) mean that an activated normal train service $ {\rm{u}\in  \mathbb{U}}$ turns around if it conflicts with the spatio-temporal disruption area, and then its turnaround train service ${\rm{v}\in \mathbb{\tilde{U}}}$ is activated. 
Constraints (\ref{turnaround3})-(\ref{turnaround4}) mean that the arrival/departure times of an activated train service ${\rm{v}\in \mathbb{\tilde{U}}}$ at station ${\rm{r}\in \mathbb{R}}$ remain unchanged if it is activated. On the contrary, constraints (\ref{turnaround5})-(\ref{turnaround6})) mean that the arrival/departure times of an activated train service ${\rm{v}\in \mathbb{\tilde{U}}}$ at station ${\rm{r}\in \mathbb{R}}$ euqal -1 if it is not activated. Remark that the station arrival/departure time of train service ${\rm{v}\in \mathbb{\tilde{U}}}$ is calculated in advance, which is an input to our model. Constraint (\ref{turnaround7}) means that a train service ${\rm{u}\in \mathbb{U}}$ does not turn around if it is not activated, and also a train service ${\rm{u}\in \mathbb{U}}$ does not turn around if it does not conflict with the spatio-temporal area. The turnaround train service ${\rm{v}\in \mathbb{\tilde{U}}}$ of a train service ${\rm{u}\in \mathbb{U}}$ would not be activated in both two mentioned conditions. 
\subsection{Identification of visited stations}
Station identification refers to the stations visited by a train service. We use binary variable $s_{{\rm{u,r}}}$ to denote whethere train service u visit station r. It is determined by

\begin{subequations}
\begin{equation}
 s_{{\rm{u,r}}}\ge a_{{\rm{u}}}-\Theta_{{\rm{u}}}, \forall  {\rm{u} \in \mathbb{U}}, {\rm{r}\in  \mathbb{R}},\label{identification1}\qquad\qquad\qquad\qquad\qquad\
\end{equation}
\begin{equation}
 s_{{\rm{u,r}}}\le a_{{\rm{u}}}, \forall  {\rm{u} \in \mathbb{U}}\cup\mathbb{\tilde{U}}, {\rm{r}\in  \mathbb{R}},\label{identification2}\qquad\qquad\qquad\qquad\qquad\
\end{equation}
\end{subequations}
\begin{subequations}
\begin{equation}
 s_{{\rm{u,r}}}\ge1+(a_{{\rm{u}}}+\Theta_{{\rm{u}}}+{\rm{f}}_{{\rm{u}}}-3), \forall  {\rm{u} \in \mathbb{U}}, {\rm{r}\in \mathbb{R}_{\rm{O}}^{1}},\label{identification3}\qquad\qquad
\end{equation}
\begin{equation}
s_{{\rm{u,r}}}\le(3-a_{{\rm{u}}}-\Theta_{{\rm{u}}}-{\rm{f}}_{{\rm{u}}}), \forall  {\rm{u} \in \mathbb{U}}, {\rm{r}\in \mathbb{R}\setminus \mathbb{R}_{\rm{O}}^{1}},\label{identification4}\qquad\qquad
\end{equation}
\end{subequations}
\begin{subequations}
\begin{equation}
\quad \ s_{{\rm{u,r}}}\ge1+(a_{{\rm{u}}}+\Theta_{{\rm{u}}}-{\rm{f}}_{{\rm{u}}}-2), \forall  {\rm{u} \in \mathbb{U}}, {\rm{r}\in \mathbb{R}_{\rm{O}}^{2}},\label{identification5}\qquad\qquad\quad
\end{equation}
\begin{equation}
\quad \ s_{{\rm{u,r}}}\le(2-a_{{\rm{u}}}-\Theta_{{\rm{u}}}+{\rm{f}}_{{\rm{u}}}), \forall  {\rm{u} \in \mathbb{U}}, {\rm{r}\in \mathbb{R}\setminus \mathbb{R}_{\rm{O}}^{2}},\label{identification6}\qquad\qquad\quad
\end{equation}
\end{subequations}
\begin{subequations}
\begin{equation}
\qquad \qquad s_{{\rm{v,r^{+}}}}\ge1+(a_{{\rm{v}}}+\sum_{{\rm{u}\in \mathbb{U}}}\delta_{{\rm{u,v}}}^{{\rm{r}}}+{\rm{f}}_{{\rm{v}}}-3), \forall {\rm{v} \in \mathbb{\tilde{U}}}, {\rm{r^{+}}\in \{r,...,| \mathbb{R}|\}},{\rm{r}\in \mathbb{R}},\label{identification7}
\end{equation}
\begin{equation}
\qquad \quad s_{{\rm{v,r^{+}}}}\le(3-a_{{\rm{v}}}-\sum_{{\rm{u}\in \mathbb{U}}}\delta_{{\rm{u,v}}}^{{\rm{r}}}-{\rm{f}}_{{\rm{v}}}), \forall {\rm{v} \in \mathbb{\tilde{U}}}, {\rm{r^{+}}\in \{1,...,r-1\}},{\rm{r}\in \mathbb{R}},\label{identification8}
\end{equation}
\end{subequations}
\begin{subequations}
\begin{equation}
\quad\qquad \ s_{{\rm{v,r^{-}}}}\ge1+(a_{{\rm{v}}}+\sum_{{\rm{u}\in \mathbb{U}}}\delta_{{\rm{u,v}}}^{{\rm{r}}}-{\rm{f}}_{{\rm{v}}}-2), \forall {\rm{v} \in \mathbb{\tilde{U}}}, {\rm{r^{-}}\in \{1,...,r\}},{\rm{r}\in \mathbb{R}},\label{identification9}
\end{equation}
\begin{equation}
\quad\qquad \ s_{{\rm{v,r^{-}}}}\le(2-a_{{\rm{v}}}-\sum_{{\rm{u}\in \mathbb{U}}}\delta_{{\rm{u,v}}}^{{\rm{r}}}+{\rm{f}}_{{\rm{v}}}), \forall {\rm{v} \in \mathbb{\tilde{U}}}, {\rm{r^{-}}\in \{r+1,...,| \mathbb{R}|\}},{\rm{r}\in \mathbb{R}}.\label{identification10}
\end{equation}
\end{subequations}
Constraint (\ref{identification1}) means that all the stations are visited by an activated train service ${\rm{u} \in \mathbb{U}}$ if it does not conflict with the spatio-temporal disruption area, and constraint (\ref{identification2}) means that all the stations are not visited by a train service ${\rm{u} \in \mathbb{U}}$ if it is not activated. 
Constraint (\ref{identification3}) means that all the stations in MS-1 are visited by an activated positive train service ${\rm{u} \in \mathbb{U}}$ if it conflicts with the spatio-temporal disruption area, and constraint (\ref{identification4}) means that an activated positive train service ${\rm{u} \in \mathbb{U}}$ would not visit station ${\rm{r}\in \mathbb{R}\setminus \mathbb{R}_{\rm{O}}^{1}}$ if it conflicts with the spatio-temporal disruption area. Constraint (\ref{identification5}) means that that all the stations in MS-2 are visited by an activated negative train service ${\rm{u} \in \mathbb{U}}$ if train service u conflicts with the spatio-temporal disruption area, and constraint (\ref{identification6}) means that an activated positive train service ${\rm{u} \in \mathbb{U}}$ would not visit station ${\rm{r}\in \mathbb{R}\setminus \mathbb{R}_{\rm{O}}^{2}}$ if it conflicts with the spatio-temporal disruption area.

For an activated positive turnaround train service, constraint (\ref{identification7}) means that the train service ${\rm{v} \in \mathbb{\tilde{U}}}$ visits station ${\rm{r^{+}}\in \{r,...,| \mathbb{R}|\}}$ if it is activated at station r$\in \mathbb{R}$, and constraint (\ref{identification8}) means that the train service ${\rm{v} \in \mathbb{\tilde{U}}}$ does not visit station ${\rm{r^{+}}\in \{1,...,r-1\}}$ if it is activated at station r$\in \mathbb{R}$. Similarly, for an activated negative train service, constraint (\ref{identification9}) means that the train service ${\rm{v} \in \mathbb{\tilde{U}}}$ visits station ${\rm{r^{-}}\in \{1,...,r\}}$ if it is activated at station r$\in \mathbb{R}$, and constraint (\ref{identification10}) means that the train service ${\rm{v} \in \mathbb{\tilde{U}}}$ does not visit station ${\rm{r^{-}}\in \{r+1,...,| \mathbb{R}|\}}$ if it is activated at station r$\in \mathbb{R}$.
\subsection{Headway constraints}
Headway constraints are considered in metro timetabling to ensure the safety distance between two successive trains. Train headway is typically classified into four types: arrival-arrival (AA), arrival-departure (AD), departure-arrival (DA), and departure-departure (DD). The headway constraints can be illustrated as follows:

\begin{equation}
\quad \ s_{{\rm{u,r}}}+s_{{\rm{v,r}}}\le 2-\theta_{{\rm{u,v}}}^{{\rm{r}}}, \forall  {\rm{u} \in \mathbb{U}, {\rm{v}} \in \mathbb{\tilde{U}}}, {\rm{r}\in  \mathbb{R}},\label{headway}
\end{equation}
where binary parameter $\theta_{{\rm{u,v}}}^{{\rm{r}}}$ with value 1 indicating train service ${\rm{u} \in \mathbb{U}}$ and train service ${\rm{v}} \in \mathbb{\tilde{U}}$ meet the headway constraints at station ${\rm{r}\in  \mathbb{R}}$. Constraint(\ref{headway}) means that train service ${\rm{u} \in \mathbb{U}}$ and train service $\mathbb{\tilde{U}}$ can not visit station ${\rm{r}\in  \mathbb{R}}$ if they do not satisfy the minimum headway. Remark that we only need to restrict the relation between train services in set $\mathbb{U}$ and train services in set $\mathbb{\tilde{U}}$, while other conditions follow the the minimum headway.
\subsection{Passenger assignment}
Recall that passengers are either assigned to train services or fail to board any train. Passenger assignment constraints can be shown as

\begin{subequations}
\begin{equation}
\quad \ \sum_{{\rm{u}}\in \mathbb{U}}x_{{\rm{p}}}^{{\rm{u}}}+x_{{\rm{p}}}={\rm{\hat{n}_{p}}}, \forall {\rm{p}\in  \mathbb{P}},\label{passenger_assignment_1}\qquad\qquad\qquad
\end{equation}
\begin{equation}
\quad \ x_{{\rm{p}}}^{{\rm{u}}} \le {\rm{\hat{n}_{p}}}\cdot \left( \begin{matrix} t_{{\rm{u,\hat{o}_{p}}}}^{d}+1\\ {\tilde{\rm{f}}_{\rm{u,p}}}\\ {\rm{\tilde{w}_{u,p}}}\\a_{{\rm{u}}} \end{matrix} \right), \forall {\rm{u}}\in \mathbb{U}\cup\mathbb{\tilde{U}}, {\rm{p}}\in \mathbb{P}.\label{passenger_assignment_2}
\end{equation}
\end{subequations}
Constraint (\ref{passenger_assignment_1}) denotes the flow conservation of passenger assignment. Constraint(\ref{passenger_assignment_2}) means that the passenger flow p can be assigned into train service u only if: (a) the train service ${\rm{u}}\in \mathbb{U}\cup\mathbb{\tilde{U}}$ departures from the origin station of passenger flow p; (b) the direction of train service ${\rm{u}}\in \mathbb{U}\cup\mathbb{\tilde{U}}$ is same as the direction of passenger group ${\rm{p}}\in \mathbb{P}$; (c) the arrival time of train service ${\rm{u}}\in \mathbb{U}\cup\mathbb{\tilde{U}}$ at the origin station of passenger flow ${\rm{p}}\in \mathbb{P}$ is not less than the production time of passenger group ${\rm{p}}\in \mathbb{P}$; (d) the train service ${\rm{u}}\in \mathbb{U}\cup\mathbb{\tilde{U}}$ is activated. The passenger flow ${\rm{p}}\in \mathbb{P}$ can not be assigned into train service ${\rm{u}}\in \mathbb{U}\cup\mathbb{\tilde{U}}$ if either of the condition (a)-(d) can not be met.
\subsection{Train capacity}
Recall that the number of passengers on the train can not exceed the train capacity, which is denoted as

\begin{subequations}
\begin{equation}
\ \quad C_{{\rm{u}}}^{{\rm{r}}}=\sum_{{\rm{p}\in  \mathbb{P}}}x_{{\rm{p}}}^{{\rm{u}}}\varphi_{{\rm{p}}}^{{\rm{u,r}}}, \forall {\rm{u}\in  \mathbb{U}}\cup\mathbb{\tilde{U}},\rm{r}\in \mathbb{R},\label{capacity1_1}\qquad
\end{equation}
\begin{equation}
\ \quad C_{{\rm{u}}}^{{\rm{r}}} \le {\rm{\bar{C}_{u}}},  \forall {\rm{u}\in  \mathbb{U}}\cup\mathbb{\tilde{U}},\rm{r}\in \mathbb{R},\label{capacity1_2}\qquad\qquad\quad
\end{equation}
\end{subequations}
where the binary parameter $\varphi_{{\rm{p}}}^{{\rm{u,r}}}$ with value 1 indicating the passenger group ${\rm{p}}\in \mathbb{P}$ is onboard after it arrives at station $\rm{r}\in \mathbb{R}$ by taking train service ${\rm{u}}\in \mathbb{U}\cup\mathbb{\tilde{U}}$, and 0 otherwise. A tailored dynamic filling algorithm is proposed to determine the value of $\varphi_{{\rm{p}}}^{{\rm{u,r}}}$  (see as \ref{algorithm_3}). Constraint (\ref{capacity1_1}) presents the formulation on the number of passengers on the train after it arrives at each station. Constraint (\ref{capacity1_2}) denotes that the number of passengers in the train can not exceed train capacity. Note that this formulation can also deal with the condition while trains have different transport capacities.
\begin{figure}
		\centering 
		\begin{minipage}{.49\linewidth}
		\includegraphics[width=0.9\textwidth]{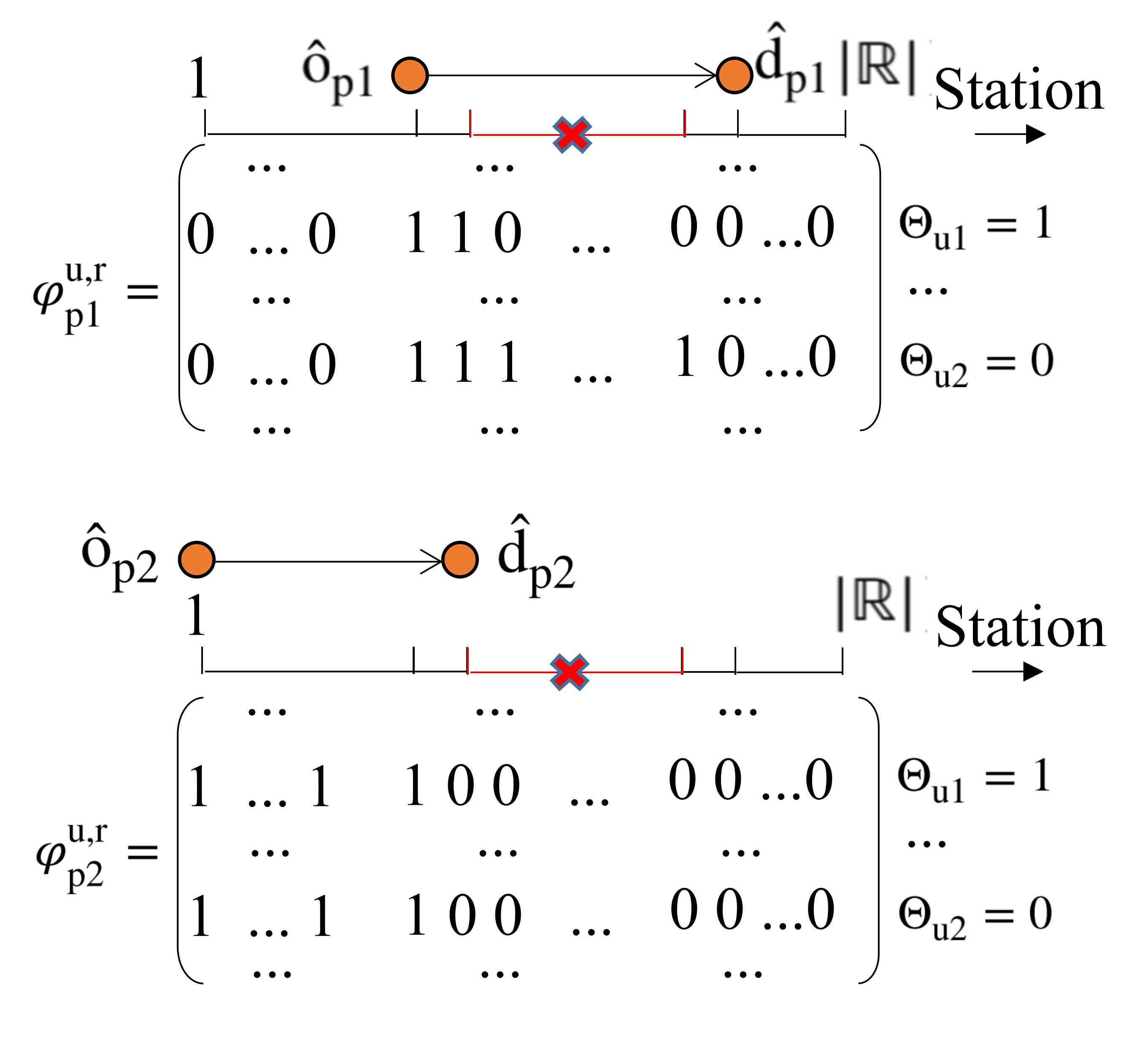}
		\caption{Identification of passenger holding}\label{train_capacity_parameter}
		\end{minipage}
		\begin{minipage}{.49\linewidth}
		\centering 
		\includegraphics[width=0.9\textwidth]{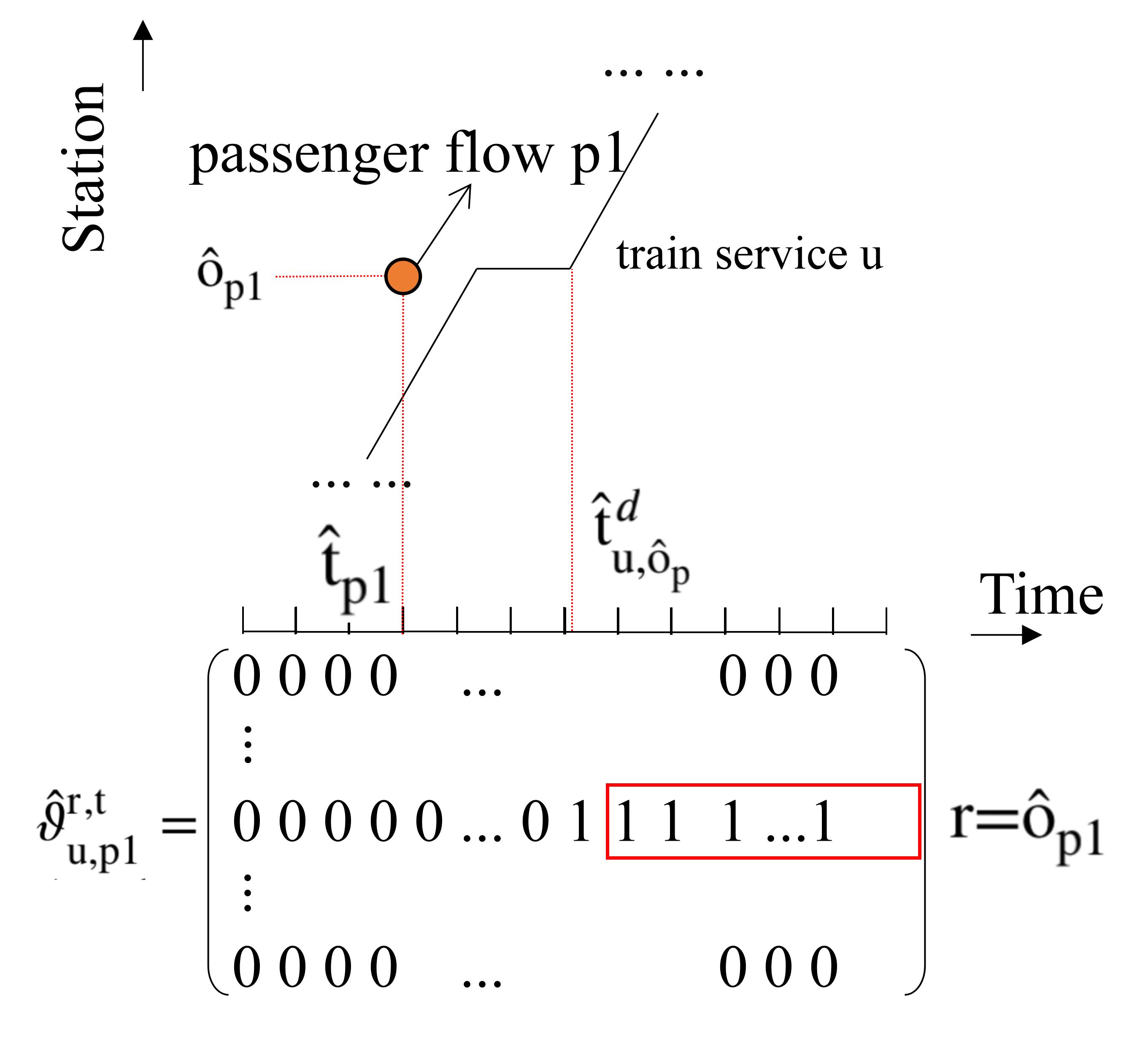}
		\caption{Identification of departure passenger flow}\label{accumulating_departure}
		\end{minipage}
\end{figure}

Figure \ref{train_capacity_parameter} depicts an illustration of the parameter $\varphi_{{\rm{p}}}^{{\rm{u,r}}}$. The passenger flows are classified into two types: those that travel through the disruption area, denoted as p1, and those that do not travel through the disruption area, denoted as p2. For the passenger flow p1, if this passenger flow has boarded train service u1 (train service u1 conflicts with spatio-temporal area, i.e., $\Theta_{{\rm{u1}}}=1$), then the parameter $\varphi_{{\rm{p1}}}^{{\rm{u1,r}}}$ equals zero from station ${\rm{\hat{o}_{p1}}}$ to the terminal station of disruption area; meanwhile, if this passenger flow has boarded train service u2 (train service u2 conflicts with spatio ; For the passenger flow p2, regardless of whatever train service is used, the parameter $\varphi_{{\rm{p2}}}^{{\rm{u,r}}}$ equals zero from station ${\rm{\hat{o}_{p2}}} $ to ${\rm{\hat{d}_{p2}}}-1$.
\subsection{Measurement of passenger accumulation}
Passenger accumulation is an indicator for assessing the quality of an operational timetable. Passenger accumulation is often assessed by stations, and passenger flow associated with a station is classified into three types: arrival flow, departure flow, and accumulation flow. The arrival flow refers to passengers arriving at a station, and the departure flow refers to people departing from this station. The accumulation flow represents passengers waiting at a station, and this sort of passenger flow indicates the station's pressure. The metro managers can first acquire real-time passenger demand by observing real-time passenger accumulation in each station, and then response strategies are implemented. Second, we should pay closer attention to passenger accumulation at the disruption area's terminal stations during the disruption time, as passenger accumulation in these stations grows quickly, posing safety threats to a station.

The idea of instantaneous accumulation flow is introduced to describe the instantaneous variation of passengers at a disruption area's terminal station, and the formulation of instantaneous passenger flow is shown below:

\begin{subequations}
\begin{equation}
\ \quad \mathscr{G}_{{\rm{p}}}^{{\rm{r,t}}}=\hat{{\rm{f}}}_{{\rm{p}}}{\rm{\hat{n}_{p}}}\tilde{\theta}_{{\rm{p}}}^{{\rm{t,r}}}+ \sum\limits_{{\rm{u}}\in \mathbb{U}} \check{\theta}_{{\rm{p,u}}}^{{\rm{t,r}}}x_{{\rm{p}}}^{{\rm{u}}},\forall {\rm{p}} \in \mathbb{P}, {\rm{r}} \in \mathbb{R}_{\rm{T}}^{1}, {\rm{t}} \in  \mathbb{T}, \label{accumulation1_1}
\end{equation}
\begin{equation}
\qquad \qquad \mathscr{G}_{{\rm{p}}}^{{\rm{r,t}}}=(1-\hat{{\rm{f}}}_{{\rm{p}}}){\rm{\hat{n}_{p}}}\tilde{\theta}_{{\rm{p}}}^{{\rm{t,r}}}+ \sum\limits_{{\rm{u}}\in \mathbb{U}} \check{\theta}_{{\rm{p,u}}}^{{\rm{t,r}}}x_{{\rm{p}}}^{{\rm{u}}},\forall {\rm{p}} \in \mathbb{P}, {\rm{r}} \in \mathbb{R}_{\rm{T}}^{2}, {\rm{t}} \in  \mathbb{T}, \label{accumulation1_2}
\end{equation}
\end{subequations}
where binary parameter $\tilde{\theta}_{{\rm{p}}}^{{\rm{t,r}}}$ with value 1 indicating that passenger group ${\rm{p}}\in \mathbb{P}$ originates from station ${\rm{r}} \in  \mathbb{R}$ at time ${\rm{t}} \in  \mathbb{T}$, and 0 otherwise. Binary parameter $\check{\theta}_{{\rm{p,u}}}^{{\rm{t,r}}}$ with value 1 indicating passenger group ${\rm{p}}\in \mathbb{P}$ who takes train service ${\rm{u}}\in \mathbb{U}\cup\mathbb{\tilde{U}}$ demands transferring at station ${\rm{r}} \in  \mathbb{R}$ at time ${\rm{t}} \in  \mathbb{T}$, and 0 otherwise. Constraints (\ref{accumulation1_1})-(\ref{accumulation1_2}) present the formulation of instantaneous accumulation flow at station $\mathbb{R}_{\rm{T}}^{1}$ and $ \mathbb{R}_{\rm{T}}^{2}$ separately. The instantaneous accumulation flow in time ${\rm{t}} \in  \mathbb{T}$ at terminal station ${\rm{r}} \in \mathbb{R}_{\rm{T}}^{1}\cup \mathbb{R}_{\rm{T}}^{2}$ is compromised of two parts: the one is the passenger flow originating from station ${\rm{r}} \in \mathbb{R}_{\rm{T}}^{1}\cup \mathbb{R}_{\rm{T}}^{2}$ at time ${\rm{t}} \in  \mathbb{T}$; the other is the passenger flow transferring at station ${\rm{r}} \in \mathbb{R}_{\rm{T}}^{1}\cup \mathbb{R}_{\rm{T}}^{2}$ at time ${\rm{t}} \in  \mathbb{T}$, and transfer passenger flow exists only at the terminal station of the disruption area.  Note that, the instantaneous accumulation flow at the terminal station of disruption area would be an input for the design of response vehicles, as the accumulated passengers have to transfer at this terminal station.\\
\textbf{Property 1. (uniqueness of instantaneous arrival parameter)} the instantaneous arrival parameters $\tilde{\theta}_{{\rm{p}}}^{{\rm{t,r}}}$ have following properties:

 \begin{equation}
 \sum\limits_{{\rm{r}}\in \mathbb{R}_{{\rm{O}}}}\sum\limits_{{\rm{t}}\in \mathbb{T}} \tilde{\theta}_{{\rm{p}}}^{{\rm{t,r}}}=1, \forall {\rm{p}} \in \mathbb{P}, \label{property1}\qquad\qquad\qquad\quad
  \end{equation}
 \textbf{Property 2.  (weak uniqueness of instantaneous transfer parameter)} the instantaneous arrival parameters $\check{\theta}_{{\rm{p,u}}}^{{\rm{t,r}}}$ have following properties:

  \begin{equation}
 \sum\limits_{{\rm{r}}\in \mathbb{R}_{{\rm{O}}}}\sum\limits_{{\rm{t}}\in \mathbb{T}} \sum\limits_{{\rm{u}}\in \mathbb{U}}\check{\theta}_{{\rm{p,u}}}^{{\rm{t,r}}} \le1, \forall {\rm{p}} \in \mathbb{P}.\qquad\qquad\quad
  \end{equation}
 
The property 1-2 can be proved by their definitions. The weak uniqueness denotes that $\sum\limits_{{\rm{r}}\in \mathbb{R}_{{\rm{O}}}}\sum\limits_{{\rm{t}}\in \mathbb{T}} \sum\limits_{{\rm{u}}\in \mathbb{U}}\check{\theta}_{{\rm{p,u}}}^{{\rm{t,r}}} =1$ only if $\sum\limits_{{\rm{r}}\in \mathbb{R}_{{\rm{T}}}}\varphi_{{\rm{p}}}^{{\rm{r}}}=1$; and $\sum\limits_{{\rm{r}}\in \mathbb{R}_{{\rm{O}}}}\sum\limits_{{\rm{t}}\in \mathbb{T}} \sum\limits_{{\rm{u}}\in \mathbb{U}}\check{\theta}_{{\rm{p,u}}}^{{\rm{t,r}}} =0$ otherwise.

The formulation of instantaneous passenger flow can be converted into its accumulated form by replacing the instantaneous parameters. Passenger accumulation can be measured as follows:

\begin{subequations}
\begin{equation}
\qquad \quad\qquad A_{{\rm{p}}}^{{\rm{r,t}}}=\begin{cases}{\rm{\hat{n}_{p}}}\tilde{\vartheta}_{{\rm{p}}}^{{\rm{t,r}}}+ \sum\limits_{{\rm{u}}\in \mathbb{U}\cup\mathbb{\tilde{U}}}\check{\vartheta}_{{\rm{p,u}}}^{{\rm{t,r}}}x_{{\rm{p}}}^{{\rm{u}}},\forall {\rm{p}} \in \mathbb{P}, {\rm{r}} \in \mathbb{R}_{\rm{O}}, {\rm{t}} \in  \mathbb{T}; \\ 0, \forall {\rm{p}} \in \mathbb{P}, {\rm{r}} \in \mathbb{R} \setminus \mathbb{R}_{\rm{O}}, {\rm{t}} \in  \mathbb{T}, \end{cases}\label{accumulation1}
\end{equation}
\begin{equation}
\ \qquad D_{{\rm{p}}}^{{\rm{r,t}}}=\begin{cases}\sum\limits_{{\rm{u}}\in \mathbb{U}\cup\mathbb{\tilde{U}}}x_{{\rm{p}}}^{{\rm{u}}}\hat{\vartheta}_{{\rm{u,p}}}^{{\rm{r,t}}},\forall {\rm{p}} \in \mathbb{P}, {\rm{r}} \in \mathbb{R}_{\rm{O}}, {\rm{t}} \in  \mathbb{T}; \\ 0, \forall {\rm{p}} \in \mathbb{P}, {\rm{r}} \in \mathbb{R} \setminus \mathbb{R}_{\rm{O}}, {\rm{t}} \in  \mathbb{T}, \end{cases}\label{accumulation2}
\end{equation}
\begin{equation}
G_{{\rm{p}}}^{{\rm{r,t}}}=A_{{\rm{p}}}^{{\rm{r,t}}}-D_{{\rm{p}}}^{{\rm{r,t}}},\forall {\rm{p}} \in \mathbb{P}, {\rm{r}} \in \mathbb{R}, {\rm{t}} \in  \mathbb{T}, \label{accumulation3}\quad
\end{equation}
\end{subequations}
where binary parameter $\tilde{\vartheta}_{{\rm{p}}}^{{\rm{t,r}}}$ with value 1 indicating that passenger group ${\rm{p}} \in \mathbb{P}$ has arrived at station ${\rm{r}} \in \mathbb{R}$ at time ${\rm{t}} \in  \mathbb{T}$, and 0 otherwise. Binary parameter $\hat{\vartheta}_{{\rm{u,p}}}^{{\rm{r,t}}}$ with value 1 indicating that passenger group ${\rm{p}} \in \mathbb{P}$ who takes train service ${\rm{u}}\in \mathbb{U}\cup\mathbb{\tilde{U}}$ has departed from station ${\rm{r}} \in \mathbb{R}$ at time ${\rm{t}} \in  \mathbb{T}$, and 0 otherwise. Constraint (\ref{accumulation1}) denotes the formulation of the accumulated arrival passenger flow at a station.  Constraint (\ref{accumulation2}) denotes the formulation of the accumulated departure passenger flow at a station. Constraint(\ref{accumulation3}) denotes the formulation of the accumulated stranded passenger flow at a station.
\begin{figure}
		\centering 
		\includegraphics[width=\textwidth]{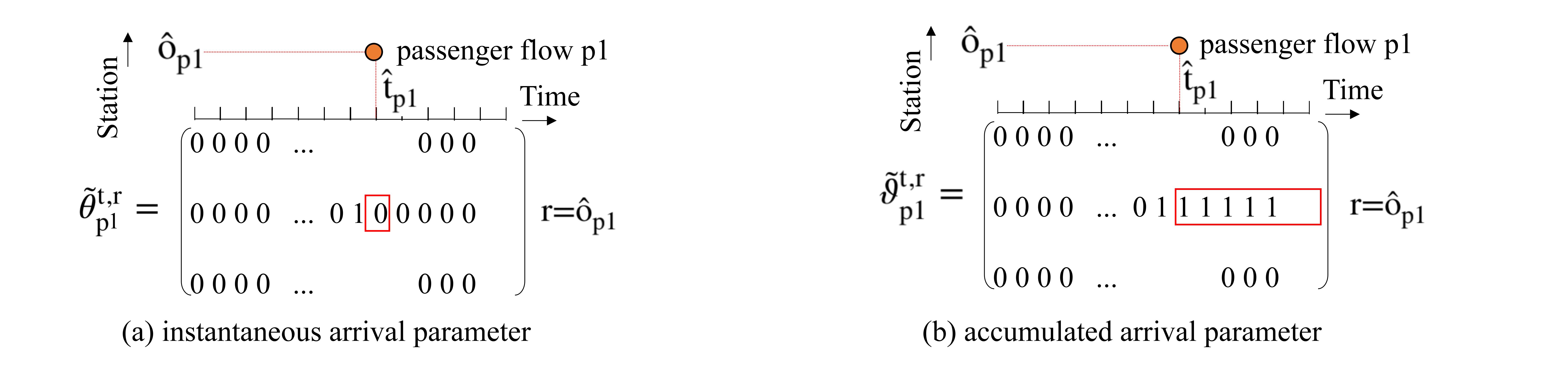}
		\caption{Identification of arrival passenger flow}\label{arrival_parameter}
	\end{figure}
	
In Figure \ref{arrival_parameter} (a)-(b), for passenger flow ${\rm{p1}\in \mathbb{P}}$, there are two parameters describing the arrival of passenger flow p1: instantaneous arrival parameter $\tilde{\theta}_{{\rm{p1}}}^{{\rm{t,r}}}$ and accumulated arrival parameter $\tilde{\vartheta}_{{\rm{p1}}}^{{\rm{t,r}}}$. For the instantaneous parameter, $\tilde{\theta}_{{\rm{p1}}}^{{\rm{t,r}}}$ equals one only if the station r and time t matches with the information of passenger flow p1.i.e., $\tilde{\vartheta}_{{\rm{p1}}}^{{\rm{t,r}}}=1$ if ${\rm{r=\hat{o}_{p1}}, t=\hat{t}_{p1}}$, and 0 otherwise; while for the accumulated parameter, $\tilde{\vartheta}_{{\rm{p1}}}^{{\rm{t,r}}}=1$ if ${\rm{r=\hat{o}_{p1}}, t\ge \hat{t}_{p1}}$, and 0 otherwise.

\begin{figure}
		\centering 
		\includegraphics[width=\textwidth]{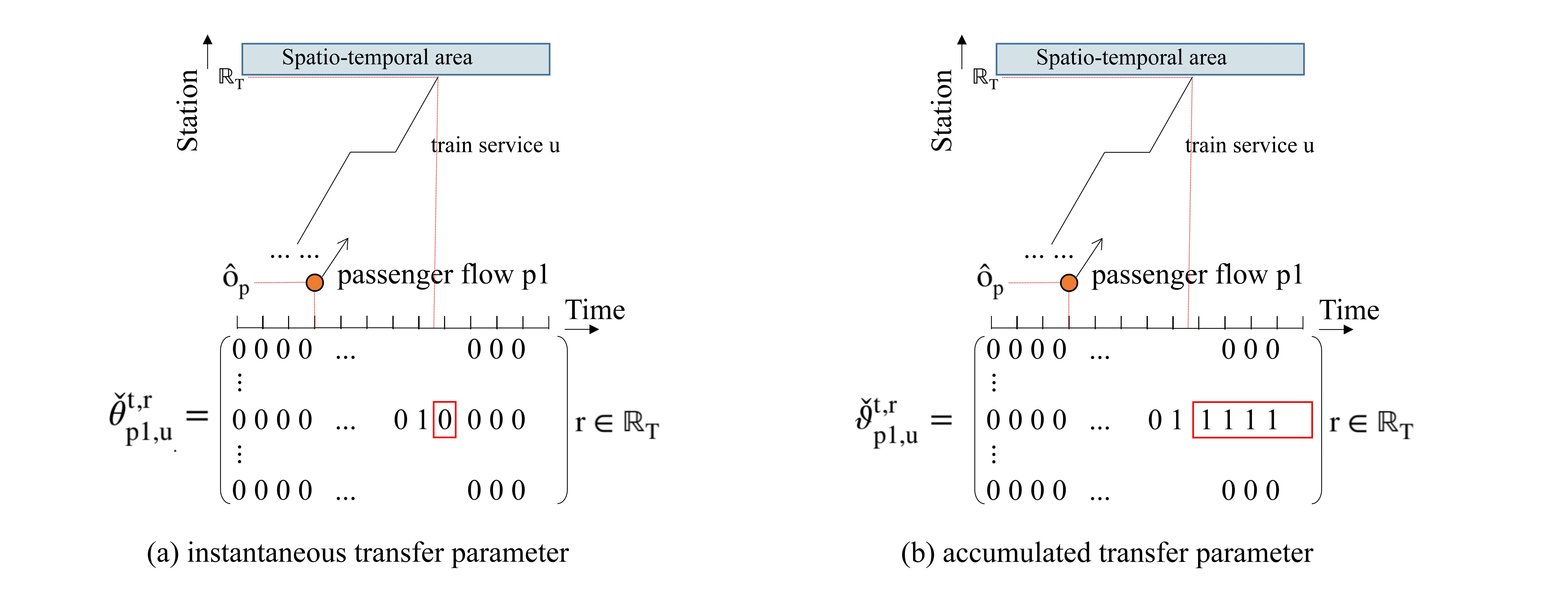}
		\caption{Identification of transfer passenger flow}\label{transfer_parameter}
	\end{figure}

In Figure \ref{transfer_parameter} (a)-(b), for the instantaneous transfer flow p1, the parameter $\check{\theta}_{{\rm{p,u}}}^{{\rm{t,r}}}$ equals one only if the passenger flow p1 arrives at his transfer station at time t.i.e., $\check{\theta}_{{\rm{p,u}}}^{{\rm{t,r}}}$=1 if t=${\rm{t}}_{{\rm{u,r}}}^{a}, \varphi_{{\rm{p}}}^{{\rm{u,r}}}=1, {\rm{r}} \in \mathbb{R}_{\rm{T}}$, and 0 otherwise; while for the accumulated transfer flow p1, the parameter $\varphi_{{\rm{p}}}^{{\rm{u,r}}}$ equals one if the passenger flow p1 has arrived at his transfer station $\mathbb{R}_{\rm{T}}$ at time t.i.e.,$\check{\theta}_{{\rm{p,u}}}^{{\rm{t,r}}}$=1 if t $\ge {\rm{t}}_{{\rm{u,r}}}^{a}, \varphi_{{\rm{p}}}^{{\rm{u,r}}}=1, {\rm{r}} \in \mathbb{R}_{\rm{T}}$, and 0 otherwise.

As shown in Figure \ref{accumulating_departure}, for the accumulated transfer flow p1 (${\rm{p1}\in \mathbb{P}}$), the parameter $\hat{\vartheta}_{{\rm{u,p1}}}^{{\rm{r,t}}}$ equals one if the passenger flow p1 has departed from station r at time t.i.e., $\hat{\vartheta}_{{\rm{u,p}}}^{{\rm{r,t}}}=1$ if t $\ge {\rm{t}}_{{\rm{u,r}}}^{d}, \varphi_{{\rm{p}}}^{{\rm{u,r}}}=1$, and 0 otherwise. \\
\textbf{Remark 2.} The instantaneous parameters and their accumulation parameters satisfy:

(1)$\tilde{\theta}_{{\rm{p}}}^{{\rm{t,r}}} \le \tilde{\vartheta}_{{\rm{p}}}^{{\rm{t,r}}}, \forall {\rm{p}} \in \mathbb{P},{\rm{r}} \in   \mathbb{R}_{\rm{O}}$;

(2)$\check{\theta}_{{\rm{p,u}}}^{{\rm{t,r}}}\le \check{\vartheta}_{{\rm{p,u}}}^{{\rm{t,r}}}, \forall {\rm{p}} \in \mathbb{P}, {\rm{u}\in \mathbb{U}}, {\rm{r}} \in  \mathbb{R}_{\rm{O}}$.\\
\textbf{Proof.} Remark 2 can be proved by their definitions.
\section{Stage 2: response vehicle scheduling}
In this section, we first map the passenger demand into the demand of response vehicles in the disruption area. This mathematical mapping is compromised of two parts: task mapping and demand mapping. Task mapping means to map the origin/destination of passengers into the origin/destination of response vehicles, while demand mapping means to map the real-time passenger demand into the real-time demand of response vehicles. 
Second, considering the real-time demand of response vehicles, we introduce the SO-DTA model ((\cite{Athanasios2000})) to guide the real-time routes of response vehicles.
\subsection{Passenger mapping}
$ $\\
\textbf{Remark 3. (task mapping)} Let a mapping  $F_{T}:(\mathbb{Z}_{+}^{| \mathbb{P}|}, \mathbb{Z}_{+}^{| \mathbb{P}|}) \to(\mathbb{Z}_{+}^{| \mathbb{P}|}, \mathbb{Z}_{+}^{| \mathbb{P}|})$ be given by $({\rm{\tilde{o}}}, {\rm{\tilde{d}}})=F_{T} ({\rm{\hat{o}}},{\rm{\hat{d}}})$. Then,

(1) ${\rm{\tilde{o}_{p}}}= \xi_{\rm{p}}^{1}[{\rm{\hat{f}_{p}}}{\rm{r}_{1}^{*}}+(1-{\rm{\hat{f}_{p}}}){\rm{r}_{2}^{*}}]+(1-\xi_{\rm{p}}^{1}){\rm{\hat{o}_{p}}}$,

(2) ${\rm{\tilde{d}_{p}}}=\xi_{\rm{p}}^{2}[(1-{\rm{\hat{f}_{p}}}){\rm{r}_{1}^{*}}+{\rm{\hat{f}_{p}}}{\rm{r}_{2}^{*}}]+(1-\xi_{\rm{p}}^{2}){\rm{\hat{d}_{p}}}$.

Binary parameter $\xi_{\rm{p}}^{1}/\xi_{\rm{p}}^{2}$ with value 1 indicating that the initial origin/destination of passenger flow ${\rm{p}} \in \mathbb{P}$ should be transferred, and 0 otherwise. ${\rm{r}_{1}^{*}}$ and ${\rm{r}_{2}^{*}}$ are the terminal stations of disruption area in the positive direction,i.e., ${\rm{r}_{1}^{*}}=\mathop{\arg\min}_{{\rm{r}} \in \mathbb{R}_{{\rm{D}}}}\{{\rm{r}}\}, {\rm{r}_{2}^{*}}=\mathop{\arg\max}_{{\rm{r}} \in \mathbb{R}_{{\rm{D}}}}\{{\rm{r}}\}$. Note that, task mapping aims to determine the transportation task of a passenger group in the disruption area. For example, for a passenger group with OD pair (1,7), and the disruption area is from station 3 to 5, then the transferred OD pair would be (3,5). The task of response vehicles is to transport this passenger flow from station 3 to station 5.\\
\textbf{Remark 4. (demand mapping)} Let a mapping $F_{D}:{\rm{R}}_{+}^{|{\rm{\mathbb{R}_{{\rm{O}}}}}|\times |{\rm{\mathbb{T}_{{\rm{D}}}}}|\times |{\rm{\mathbb{P}}}|} \to {\rm{R}}_{+}^{|{\rm{\mathbb{C}}}|\times |{\rm{\mathbb{M}}}|\times \{\rm{\mathbb{T}_{{\rm{R}}}}|}$ be given by $ {\rm{d}}=F_{D} (\mathscr{G})$. Then the demand of response vehicles is denoted as:

\begin{equation}
\qquad\qquad {\rm{d}}_{{\rm{i,m}}}^{{\rm{t}}}=\begin{cases}\lceil \sum\limits_{{\rm{r}} \in  \mathbb{R}_{{\rm{O}}}} \sum\limits_{{\rm{p}} \in \mathbb{P}}\sum\limits_{\tau \in  \mathbb{\tilde{T}}_{{\rm{R}}} ({\rm{t}})}\mathscr{G}_{{\rm{p}}}^{{\rm{r,\tau}}}\tilde{\delta}_{{\rm{i,r}}}^{{\rm{m,p}}}/ {\rm{C_{R}}}\rceil,  \forall {\rm{i}} \in \mathbb{C}, {\rm{t}} \in  \mathbb{T}_{{\rm{R}}}, {\rm{m}} \in  \mathbb{M}; \\ 0, else; \end{cases} 
\end{equation}

Binary parameter $\tilde{\delta}_{{\rm{i,r}}}^{{\rm{m,p}}}$ with value 1 indicating that the passenger flow ${\rm{p}} \in \mathbb{P}$ at station ${\rm{r}} \in \mathbb{R}$ matches with the transportation task (OD pair) of class-m response vehicles departing from cell i.i.e., $\tilde{\delta}_{{\rm{i,r}}}^{{\rm{m,p}}}=1$ if ${\rm{\tilde{o}_{p}}}={\rm{o_{m}}}, {\rm{\tilde{d}_{p}}}={\rm{d_{m}}}$, and cell i is station r; and 0 otherwise. 
In this section, the time set $\mathbb{T}$ is divided into $\mathbb{T}_{{\rm{R}}}$ periods, and $\mathbb{\tilde{T}}_{{\rm{R}}}$(t) denotes the set of discrete time of time period ${\rm{t}} \in  \mathbb{T}_{{\rm{R}}}$. For example, for a frequency-based response vehicle fleet, the set $\mathbb{\tilde{T}}_{{\rm{R}}}$(2)=$\{\omega_{\rm{F}},...,2\omega_{\rm{F}}\}$ if the departure frequency of response vehicle is $\omega_{\rm{F}}$. 

\subsection{Formulations}
To capture the real-time traffic dynamics, the constraints of response vehicle routing are formulated based on the well-known structure of SO-DTA. In our proposed model, the response vehicles are loaded by the method of cell transmission model(CTM).
Furthermore, the state transition of cells is formulated as flow conservation constraints. Considering the character of cells, the flow conservations are also classified by cell types. That is, the flow conservation constraints should be distinguished among the source cells, sink cells, and other cells. Also, the demand and supply functions are relaxed as flow propagation constraints (\cite{Athanasios2000}). In addition, other necessary constraints, such as the initial state of links and cells, and nonnegative variables are also concluded.

\begin{equation}
\min Z_{2}=\sum_{{\rm{t}} \in \mathbb{T}_{{\rm{R}}}}\sum_{{\rm{m}} \in \mathbb{M}}\sum_{{\rm{i}} \in \mathbb{C}\setminus  \mathbb{C}_{{\rm{S}}}}\alpha_{\rm{t}}y_{{\rm{i,m}}}^{{\rm{t}}}\label{obj2}\qquad\qquad\qquad\qquad\qquad\qquad\qquad\qquad\qquad\qquad\qquad
\end{equation}
\begin{subequations}
\begin{equation}
s.t. y_{{\rm{i,m}}}^{{\rm{t}}}-y_{{\rm{i,m}}}^{{\rm{t-1}}}-\sum_{{\rm{k} \in\Gamma(i)}}z_{{\rm{k,i,m}}}^{{\rm{t-1}}}+\sum_{{\rm{j} \in\Gamma^{-} (i)}}z_{{\rm{i,j,m}}}^{{\rm{t-1}}}=0, \forall {\rm{i}} \in \mathbb{C}\setminus \{ \mathbb{C}_{{\rm{S}}}, \mathbb{C}_{{\rm{R}}}\}, {\rm{t}} \in \{1,...,| \mathbb{T}_{{\rm{R}}}|\}, {\rm{m}} \in  \mathbb{M},\label{res_0}
\end{equation}
\begin{equation}
\ \quad y_{{\rm{i,m}}}^{{\rm{t}}}-y_{{\rm{i,m}}}^{{\rm{t-1}}}+z_{{\rm{k,i,m}}}^{{\rm{t-1}}}={\rm{d}}_{{\rm{i,m}}}^{{\rm{t-1}}}, \forall {\rm{i}} \in \mathbb{C}_{\rm{R}}, {\rm{t}} \in \{1,...,| \mathbb{T}_{{\rm{R}}}|\},  {\rm{m}} \in  \mathbb{M},\label{res_5_1}\qquad\qquad\qquad\qquad\qquad
\end{equation}
\begin{equation}
\ \quad \sum_{{\rm{t}} \in \mathbb{T}_{{\rm{R}}}}\sum_{{\rm{k} \in\Gamma({\rm{\mathbb{C}_{\rm{S}})}}}}z_{{\rm{k,{\rm{\mathbb{C}_{\rm{S}},m}}}}}^{{\rm{t}}}= \sum_{{\rm{t}} \in \mathbb{T}_{{\rm{R}}}}\sum_{{\rm{i} \in{\rm{\mathbb{C}_{\rm{R}}}}}}{\rm{d}}_{{\rm{i,m}}}^{{\rm{t-1}}}, \forall  {\rm{m}} \in  \mathbb{M},\label{res_5_2}\qquad\qquad\qquad\qquad\qquad\qquad\qquad\qquad
\end{equation}
\end{subequations}
\begin{subequations}
\begin{equation}
\ \quad \sum_{{\rm{j} \in\Gamma^{-} (i)}}z_{{\rm{i,j,m}}}^{{\rm{t}}}\le y_{{\rm{i,m}}}^{{\rm{t}}}, \forall {\rm{i}} \ne{\rm{j}} \in \mathbb{C}\setminus \{ \mathbb{C}_{{\rm{S}}}, \mathbb{C}_{{\rm{R}}}\}, {\rm{t}} \in  \mathbb{T}_{{\rm{R}}}, {\rm{m}} \in  \mathbb{M},\label{res_1}\qquad\qquad\qquad\qquad\qquad\quad
\end{equation}
\begin{equation}
\ \quad \sum_{{\rm{j} \in\Gamma^{-} (i)}}z_{{\rm{i,j,m}}}^{{\rm{t}}}\le {\rm{Q_{i,m}}}, \forall {\rm{i}} \in \mathbb{C}\setminus \{ \mathbb{C}_{{\rm{S}}}, \mathbb{C}_{{\rm{R}}}\}, {\rm{t}} \in  \mathbb{T}_{{\rm{R}}}, {\rm{m}} \in  \mathbb{M},\label{res_2}\qquad\qquad\qquad\qquad\qquad\qquad
\end{equation}
\end{subequations}
\begin{subequations}
\begin{equation}
\ \quad \sum_{{\rm{k} \in\Gamma(i)}}z_{{\rm{k,i,m}}}^{{\rm{t}}}\le {\rm{Q_{i,m}}}, \forall {\rm{i}} \in \mathbb{C}\setminus \{ \mathbb{C}_{{\rm{S}}}, \mathbb{C}_{{\rm{R}}}\}, {\rm{t}} \in  \mathbb{T}_{{\rm{R}}}, {\rm{m}} \in  \mathbb{M},\label{res_3}\qquad\qquad\qquad\qquad\qquad\qquad
\end{equation}
\begin{equation}
\ \quad \sum_{{\rm{k} \in\Gamma(i)}}z_{{\rm{k,i,m}}}^{{\rm{t}}}\le \frac{{\rm{w_{m}}}}{{\rm{v_{m}}}} ({\rm{N_{i,m}^{t}}}-y_{{\rm{i,m}}}^{{\rm{t}}}), \forall {\rm{i}} \in \mathbb{C}\setminus \{ \mathbb{C}_{{\rm{S}}}, \mathbb{C}_{{\rm{R}}}\}, {\rm{t}} \in  \mathbb{T}_{{\rm{R}}}, {\rm{m}} \in  \mathbb{M},\label{res_4}\qquad\qquad\qquad\quad
\end{equation}
\end{subequations}
\begin{equation}
\ \quad {\rm{d}}_{{\rm{i,m}}}^{{\rm{t}}}=\begin{cases}\lceil \sum\limits_{{\rm{r}} \in  \mathbb{R}_{{\rm{O}}}} \sum\limits_{{\rm{p}} \in \mathbb{P}}\sum\limits_{\tau \in  \mathbb{\tilde{T}}_{{\rm{R}}} ({\rm{t}})}\mathscr{G}_{{\rm{p}}}^{{\rm{r,\tau}}}\tilde{\delta}_{{\rm{i,r}}}^{{\rm{m,p}}}/ {\rm{C_{R}}}\rceil,  \forall {\rm{i}} \in \mathbb{C}, {\rm{t}} \in  \mathbb{T}_{{\rm{R}}}, {\rm{m}} \in  \mathbb{M}; \\ 0, else; \end{cases} \label{res_5_3}\qquad\qquad\qquad\quad
\end{equation}
\begin{subequations}
\begin{equation}
\ \quad y_{{\rm{i,m}}}^{{\rm{t}}}\le 0,  \forall {\rm{i}} \in \mathbb{C}, {\rm{t}}=0, {\rm{m}} \in  \mathbb{M},\label{res_6}\qquad\qquad\qquad\qquad\qquad\qquad\qquad\qquad\qquad\qquad
\end{equation}
\begin{equation}
\ \quad z_{{\rm{i,j,m}}}^{{\rm{t}}}\le 0,  \forall {\rm{i, j}} \in \mathbb{C}, {\rm{t}}=0, {\rm{m}} \in  \mathbb{M},\label{res_7}\qquad\qquad\qquad\qquad\qquad\qquad\qquad\qquad\qquad\quad
\end{equation}
\end{subequations}
\begin{subequations}
\begin{equation}
\ \quad y_{{\rm{i,m}}}^{{\rm{t}}} \ge 0, \forall {\rm{i}} \in \mathbb{C}, {\rm{t}} \in  \mathbb{T}_{{\rm{R}}}, {\rm{m}} \in  \mathbb{M},\label{res_8}\qquad\qquad\qquad\qquad\qquad\qquad\qquad\qquad\qquad\quad
\end{equation}
\begin{equation}
\ \quad z_{{\rm{i,j,m}}}^{{\rm{t}}}\ge 0, \forall {\rm{i}} \in \mathbb{C}, {\rm{t}} \in  \mathbb{T}_{{\rm{R}}}, {\rm{m}} \in  \mathbb{M}.\label{res_9}\qquad\qquad\qquad\qquad\qquad\qquad\qquad\qquad\qquad\quad
\end{equation}
\end{subequations}

The objective (\ref{obj2}) is to minimize the systematic total travel time of response vehicles. Constraint (\ref{res_0}) describes traffic flow conservation for all road cells, excluding source and sink cells. That is, the number of class-m response vehicles in a road cell i at time t is determined by the sum of this number at time t-1 and the number of class-m response vehicles going into this cell i at time t-1, and minus the number of class-m response vehicles leaving from cell i at time t-1. Note that, the classification of response vehicles is identified by their OD pairs. This flow conservation has revealed the temporal state transition of road cells. 
Constraint (\ref{res_5_1}) denotes the flow conservation constraint in the source cell.
Constraint (\ref{res_5_2}) denotes the flow conservation constraint in the sink cell.

Constraints (\ref{res_1})-(\ref{res_4}) have defined the process of traffic flow loading for class-m vehicles, which is based on the relaxation of the CTM. 
The maximum outflow from a road cell to its downstream road cells is limited by constraint (\ref{res_1})-(\ref{res_2}). Firstly, the practical outflow of road cell i is restricted by the number of response vehicles in cell i. Secondly, the outflow capacity of road cell i should also be taken into account. Note that, ${\rm{Q_{i,m}}}$ is a parameter reflecting the road condition of road cells.
In the meanwhile, the maximum inflow from upstream cells to a cell is described as constraints (\ref{res_3}) and (\ref{res_4}). Firstly, the practical inflow of a road cell i is restricted by the inflow capacity of this cell. Secondly, the available space for receiving response vehicles should also be considered. Note that, the available space is estimated by the spare space of this cell, as well as the free-flow speed and the backward shock wave propagation speed of response vehicles.
Constraints (\ref{res_5_3}) denotes the vehicle demand in each cell.
Constraints (\ref{res_6}) and (\ref{res_7}) denote the initial state of cells and links. That is, the discrete road network is empty before assigning the response vehicles. 
Constraints (\ref{res_8}) - (\ref{res_9}) denote the decision variables of this model, and the decision variables are both continuous and nonnegative.
\section{Case study}
\subsection{Basic experiment}
\subsubsection{parameter setting}
\begin{figure}
		\centering 
		\includegraphics[width=\textwidth]{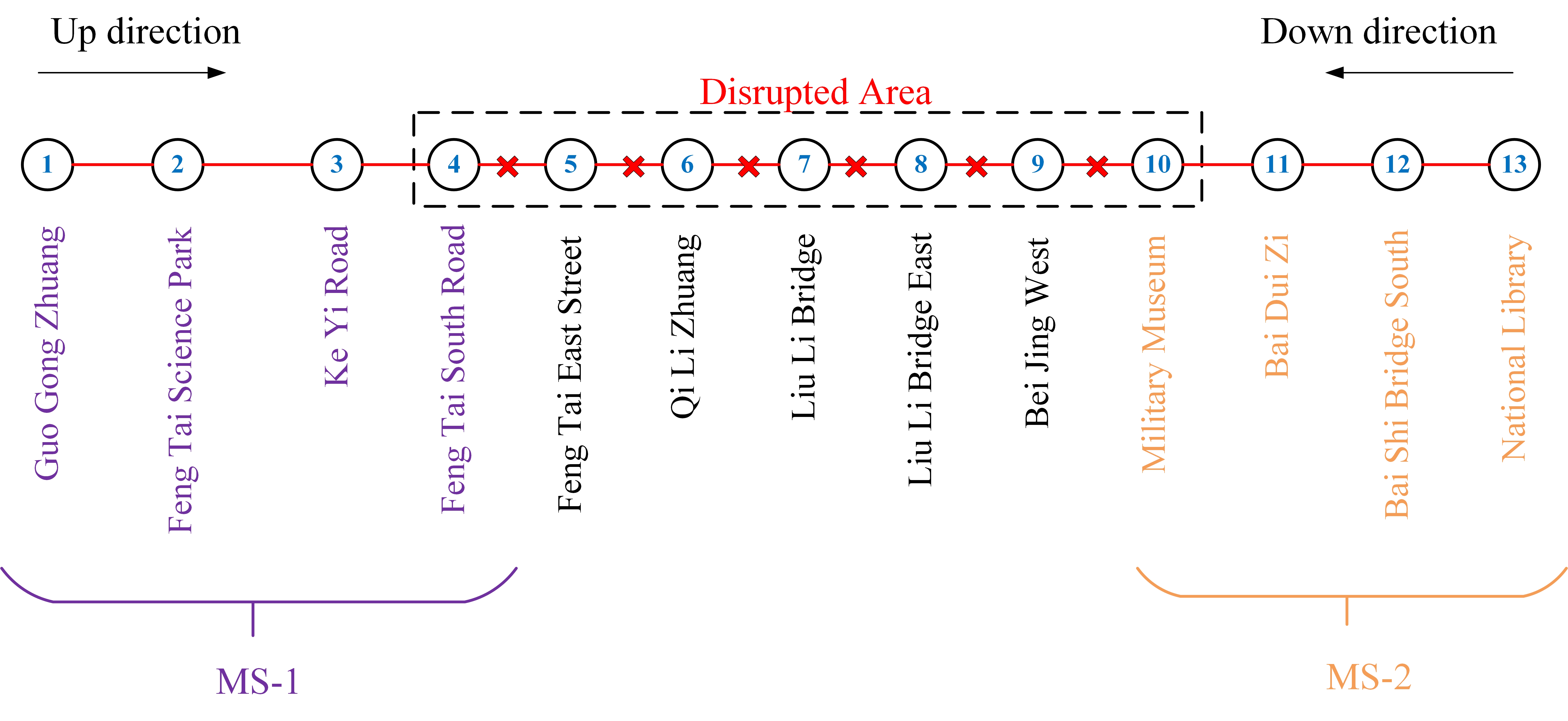}
		\caption{Beijing Metro Line 9}\label{Metro_line_9}
	\end{figure}

Figure \ref{Metro_line_9} depicts the entire Beijing Metro Line 9, which has thirteen metro stations from Guo Gong Zhuang Station (GGZ) to National Library Station (NL). An unexpected disruption is taking place between Feng Tai South Road Station (FTSR) and Military Museum Station (MM), with the disruption lasting from 8:00 to 9:00. The metro transportation system is divided into two areas: in MS-1, the station set concludes four stations: GGZ station, FTSP station, KYR station, and FTSR station. We denote this set as $\mathbb{R}_{{\rm{O}}}^{1}=\{1,2,3,4\}$; In the MS-2, it also concludes four stations: MM station, BDZ station, BSBS station, NL station, and we denote this set as $\mathbb{R}_{{\rm{O}}}^{2}=\{10,11,12,13\}$. Furthermore, the station set of disruption area can be denotes as $\mathbb{R}_{{\rm{D}}}=\{4,5,6,7,8,9,10\}$. The terminal station sets of disruption area can be denoted as: $\mathbb{R}_{{\rm{T}}}^{1}=\{4\}$, $\mathbb{R}_{{\rm{T}}}^{2}=\{10\}$.  

\begin{figure}[H]
		\centering 
		\includegraphics[width=\textwidth]{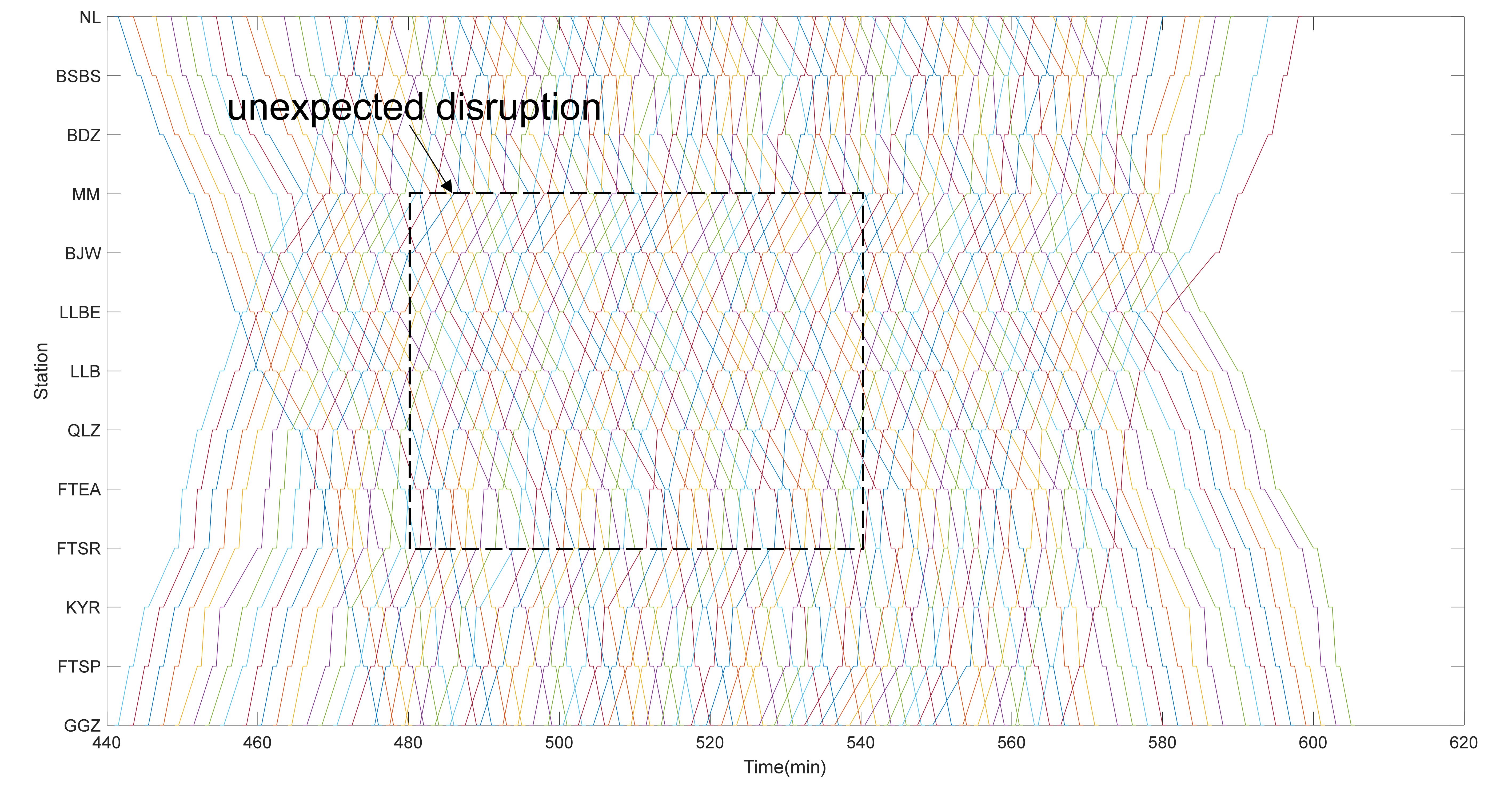}
		\caption{Normal timetable}\label{timetable1}
	\end{figure}

The partial timetable of Beijing Metro Line 9 is shown in Figure \ref{timetable1}. In the original normal timetable generated by the Yizhuang metro line's workday timetable, there are 118 train services from 7:20 to 9:30. There are 58 positive-direction train services and 60 negative-direction train services in this normal timetable. The volume of train service is set as $\rm{\bar{C}_{{\rm{u}}}}=1000$. 
\begin{table}
\centering
\caption{Parameters of response vehicles}\label{Parameters}
\resizebox{0.8\textwidth}{!}{
\begin{tabular}{ccccc}
\toprule     
Class&Length (m)&Passenger capacity&Free-flow speed (m/s)& Shock wave speed (m/s)\\ 
\midrule  
Bus&12&40&20&10\\
\bottomrule    
\end{tabular}
}
\end{table}
\begin{table}[H]
\caption{Signal positions and their fixed time plans}\label{Signal}
\resizebox{\textwidth}{!}{
\begin{tabular}{cccc}
\toprule     
Signal position (in cell \#)&Cycle time (time intervals)&Green time (time intervals)&Time when first green appear (time intervals)\\ 
\midrule  
3&5&2&0\\
4&5&2&0\\
11&5&2&4\\
20&5&2&4\\
\bottomrule    
\end{tabular}
}
\end{table}

The practical road condition should be considered when scheduling response vehicles in the disruption area. Figure \ref{road_network} of \ref{appendix_figure} depicts the topological structure of the road network along the disrupted metro line and its cell discretion grapha. The cell length is equal to the free-flow speed multiplied by the 20-second time interval. As a result, the cell length is 400 m, the maximum capacity is 11 vehicles per time step, and the maximum occupancy per cell is 33 vehicles. Because the stations from FTES to BJW are closed for safety, response vehicles evaluate twelve OD pairs: from the origin station FTSR to the destination station FTES, QLZ, LLQ, LLQE, BJW, and MM; and from the origin station MM to the destination stations BJW, LLQE, LLQ, QLZ, FTNA, and FTSR. The response vehicle parameters are consistent, as shown in Table \ref{Parameters}. The maximum flow velocity of the bus is 1992 vph for each lane, according to the triangle fundamental diagram. Table \ref{Signal} shows the signal intersections and their fixed time plans.
\subsubsection{rescheduled metro timetable}
\begin{figure}
		\centering 
		\includegraphics[width=\textwidth]{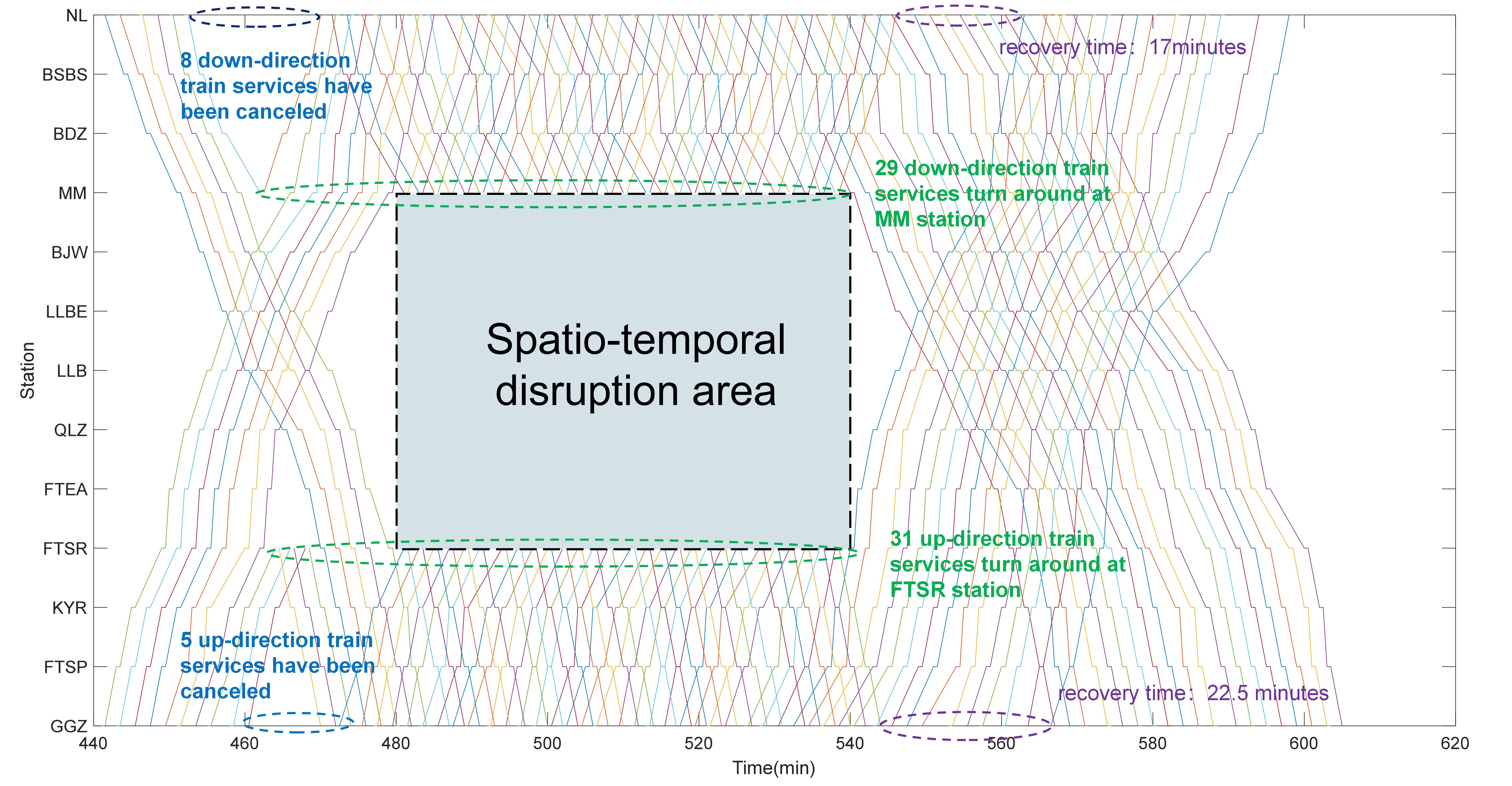}
		\caption{Rescheduled timetable}\label{timetable2}
	\end{figure}

As shown in Figure \ref{timetable1}, an unexpected disruption occurs during the normal timetable from 8:00 to 9:00, and the disruption area spans from FSTR station to MM station. The spatio-temporal disruption area is a rectangle area in the new rescheduled timetable (as shown in Figure \ref{timetable2}), and trains cannot run through it.
To ensure the safety of two successive trains traveling in the same direction, we have set all minimum headways (including AA, AD, DA, and DD headways) to one minute. 13 train services before the disruption are canceled in the rescheduled timetable to meet headway constraints with other train services. That is, if these canceled train services were activated, the headway constraints would be violated. As a result, the unexpected disruption would result in train service cancellations before it occurred.
	
To avoid the conflict between train services and the spatio-temporal disruption area, train services have to be rescheduled to turn around at terminal stations of the disruption area. As shown by Figure \ref{timetable2}, In the positive direction, 31 (of 58) train services have been rescheduled. In the meanwhile, 29 (of 60) negative direction train services have been rescheduled. These rescheduled train services are truncated as the blockage of disruption. For the positive direction train services, they have to turn around at FTSR station; while, for the negative direction train services, they turn around at MM station. Furthermore, it can be seen that no train services turn around after the disruption, and these turnaround operations are executed only before the disruption has ended. Therefore, the unexpected disruption would result in turnaround operations before the disruption has ended.

The unexpected disruption would result in a short recovery period after it has ended. During this period, the transport capacity is imbalanced. For the positive direction train services, the recovery time is 17 minutes, whereas the recovery time for negative direction train services is 22.5 minutes. The duration of the recovery period depends on the first train service after the disruption.

Statistical analysis shows that 61.9\% of train services in the normal timetable have been rescheduled or canceled, while the disruption time occupancy is 46.2\%. More effect on the normal timetable might be caused while the disruption area lies in the middle position of the normal timetable, this observation can be proved by the sensitive analysis part of this paper. As the directional difference of train services, even although the number of affected train services is similar in both two directions of train services, more disturbances would be caused for the positive direction train services as they have fewer train services. Therefore, the direction with fewer train services would occur more effect while facing with an unexpected disruption.
	
The normal metro timetable has been rescheduled to ensure the safety of trains under the disruption condition, and also the passengers are assigned to this new rescheduled timetable. Note that, passengers boarding these rescheduled train services would accumulate at the terminal stations of the disruption area, if their trip crosses the disruption area, this would bring a high safety risk to FTSR station and MM station. Therefore, response vehicles should be scheduled timely to transport these accumulated passengers.

\subsubsection{passenger measurement under the rescheduled timetable}
\begin{figure}[H]
		\centering 
		\includegraphics[width=\textwidth]{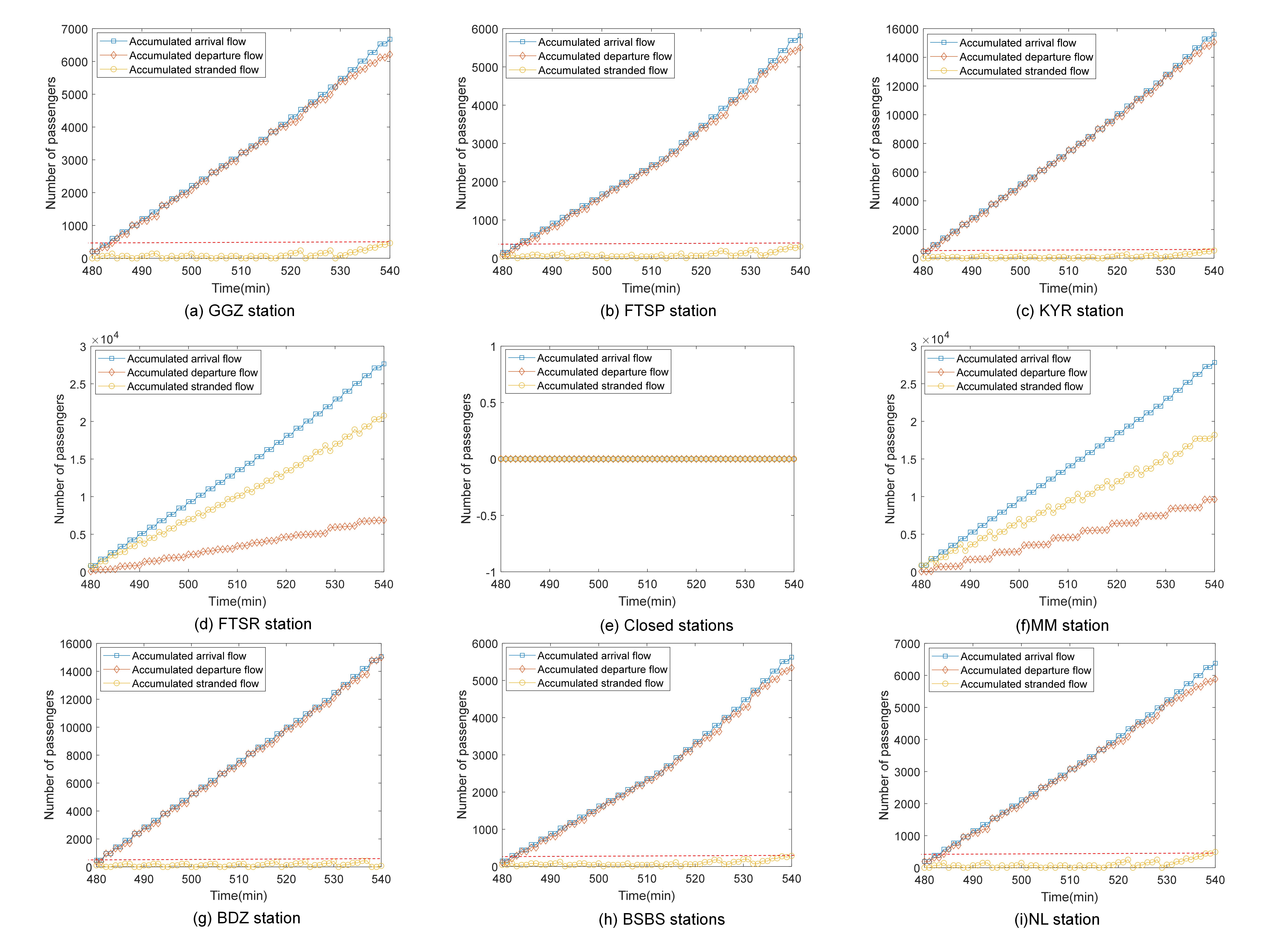}
		\caption{Passenger accumulation at each station}\label{accumulation}
\end{figure}

Passenger accumulation of each station is an important index in metro disruption management since it reflects the real-time variation of passengers in a station. By observing the real-time passenger accumulation in a station, we can assess the passenger loading pressure of each station. There are 13 metro stations in our basic experiment, and the observation time is set from 8:00 to 9:00 (disruption period).
Note that, the accumulated arrival/departure/accumulation flow at a station are calculated by the accumulated form of passenger flow,i.e., variables $\sum\limits_{{\rm{p}}\in \mathbb{P}}A_{{\rm{p}}}^{{\rm{r,t}}}/\sum\limits_{{\rm{p}}\in \mathbb{P}}D_{{\rm{p}}}^{{\rm{r,t}}}/\sum\limits_{{\rm{p}}\in \mathbb{P}}G_{{\rm{p}}}^{{\rm{r,t}}}$. The result of passenger accumulation at each station is shown in Figure \ref{accumulation}.

As shown in Figure \ref{accumulation} (a), the accumulated arrival curve approximates a linear relationship, while the accumulated departure curve climbs along the accumulated departure curve. The passengers originating from GGZ station can be boarded in time, as adequate train services are departing from GGZ station. Meanwhile, it can be seen that the passengers standed on the platform of GGZ station can be controlled under 500 all the time. In Figure \ref{accumulation} (b), the accumulated arrival curve approximates a linear relationship, and the accumulated departure curve climbs along the accumulated departure curve. Different from GGZ station, the arriving rate of passengers at FTSP station increase suddenly after 8:38, and it can also be met by our train services. The number of passengers standed on the platform of GGZ station and FTSP station can be controlled under 500 all the time, which is acceptable by metro managers.

The FTSR station, being the first terminal station of the disruption area in the positive direction, has a different accumulation tendency, as shown in Figure \ref{accumulation} (d). There are no positive direction train services departing from FTSR station; all positive direction train services finish their tasks at FTSR station, while a new negative direction train service departs from FTSR station. Thus, the newly activated turnaround train service will serve passenger demand in the negative direction; however, passengers originating from the FTSR will still have to wait in this station. Furthermore, transfer passengers wishing to transfer from the FTSR station to other stations in the positive direction must wait at the FTSR station. These two sorts of stranded passengers have caused the FTSR station's passenger accumulation to significantly expand. However, it is not worried that the response vehicles would serve the stranded passengers in time.

Figure \ref{accumulation} (e) depicts the passenger accumulation at closed stations, where the accumulation of passenger flows equals zero because these stations are closed and no passengers or trains arrive.

The passenger accumulation at MS-1 stations has been depicted in Figure \ref{accumulation} (a)-(e). Similarly, the passenger accumulation at MS-2 stations follows a similar pattern as that of MS-1 stations. The accumulated arrival curve approximates a linear relationship, as shown by the passenger accumulation at TJS station in Figure \ref{accumulation} (f), while the accumulated departure curve climbs along the accumulated arrival curve. Passengers traveling in the negative direction tend to gather at station MM. There are two sorts of passengers accumulating at MM station: passengers who originate at MM station and transfer passengers who travel from MM station to other stations. It should be noted that the response vehicles would serve the transfer passenger flows.

In Figure \ref{accumulation} (g)-(i), the passenger accumulation at stations of MS-2 (except for the terminal station of the disruption area) has been demonstrated. The accumulated arrival curve of these two stations approximates a linear relationship, while the accumulated departure curve climbs along the arrival curve. It is obvious that the passengers departing from these three stations will be picked up in time. Passenger accumulation at each station can be controlled within an acceptable range.
	\begin{figure}[H]
		\centering 
		\includegraphics[width=\textwidth]{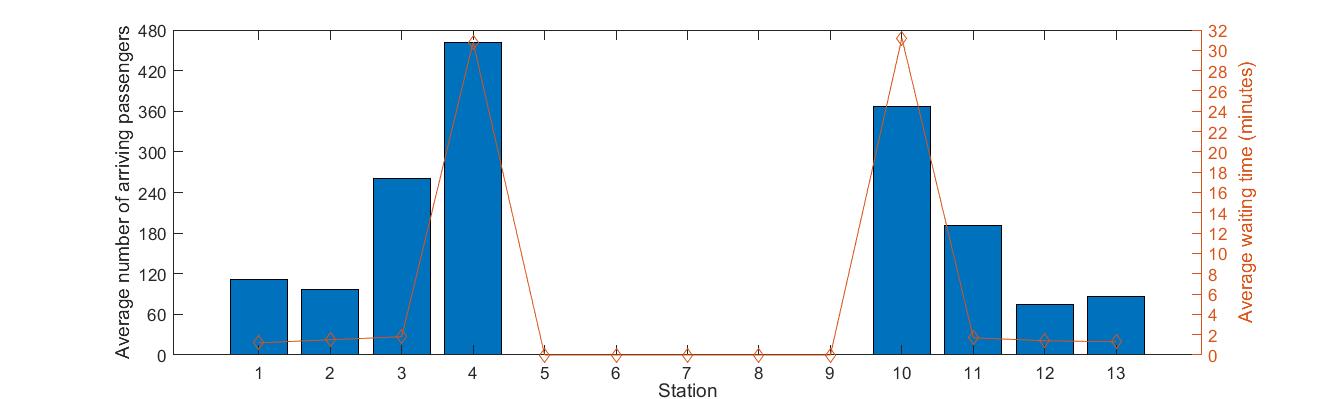}
		\caption{Average arrival number \& waiting time of passengers at each station}\label{passenger_waiting}
	\end{figure}
The average waiting time and passenger arrival are depicted in Figure \ref{passenger_waiting}. The bar chart depicts the number of passengers arriving at each stop. It can be seen that the FTSR and MM stations have the highest average number of arriving passengers. These two stations' platforms must service more than 300 people per minute. There is also no passenger demand from FTES to BJW because these five stations are blocked during the disruption period. Furthermore, the average number of people coming in MS-1 is larger than in MS-2, indicating that MS-1 is under greater passenger pressure. The line chart depicts the average time passengers spend waiting at each station. Passengers' average waiting time is greatest at FTSR and MM stations, with values greater than 30 minutes. This is because some passengers departing from these two stations must wait until the disruption is over if no response vehicles arrive. Furthermore, the average wait time at other open stations is less than 3 minutes, which is consistent with train service departure frequency. This results show that the new rescheduled timetable can provide passengers with an efficient mode of transportation, and the passengers will be picked up in time. As a result, passengers stranded at FTSR and MM stations would have a shorter wait if we offered timely response vehicle service, and other passengers would only have to wait 2 minutes before boarding a train. The impact of disruption can be mitigated by coordinating the optimization of metro and response vehicles.
	\begin{figure}[H]
		\centering 
		\includegraphics[width=\textwidth]{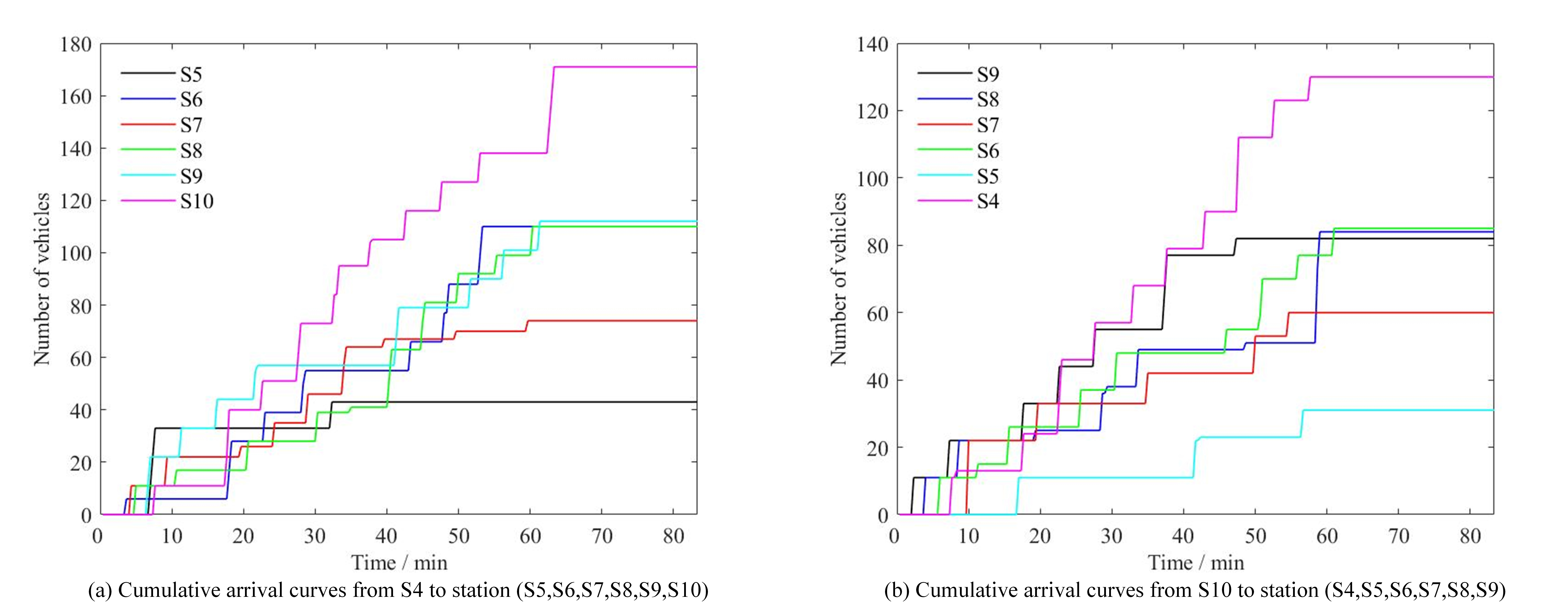}
		\caption{cumulative arrive curves of response vehicles}\label{cum_curve}
	\end{figure}
\subsubsection{scheduling of response vehicles}
Fig \ref{cum_curve} (a) and Fig \ref{cum_curve} (b) depict the cumulative arrival curves in two directions (i.e., positive direction and negative direction). The cumulative curves show a step-by-step change because the response vehicles are scheduled at fixed time intervals of 5 minutes. We can obtain the departure sequence and fleet size of response vehicles, which are combined with a space-time diagram to provide real-time guidance.
	
The network clearance time (NCT) is made up of two components: path holding time and path travel time. For example, the larger the NCT without considering dynamic traffic, the farther the station is from the origin station. However, the NCT of S4-S7 and S4-S8 is quite similar because S8 receives more response vehicles than S7. The NCT gap between S10-S5 and S10-S4 is 1 minute in the direction of S4, whereas the corresponding fleet size gap is 84 vehicles. The slope of the cumulative arrival curves at station S4 is greater becausethe shorter routes are prioritized for the fleets reaching station S4.
\begin{table}[H]
\centering
\caption{Results of fleet size and NCT for all OD pairs}\label{Results}
\resizebox{0.5\textwidth}{!}{
\begin{tabular}{ccc}
\toprule     
OD pair&Number of vehicles&NCT of response vehicles (min)\\ 
\midrule  
S4-S5&36&32.67\\
S4-S6&92&53.67\\
S4-S7&62&60.00\\
S4-S8&93&60.33\\
S4-S9&94&61.33\\
S4-S10&144&63.67\\
\midrule
S10-S9&71&47.33\\
S10-S8&72&59.33\\
S10-S7&51&55\\
S10-S6&72&61.33\\
S10-S5&26&57\\
S10-S4&110&58\\
\bottomrule    
\end{tabular}
}
\end{table}
The results of fleet size and network clearance time for various OD pairs are summarized in Table \ref{Results}. The maximum network clearance time provides a lower time limit for temporary exclusive lanes. The exclusive lanes are maintained in this case for at least 63.67 minutes, which can be used as the time threshold for designating exclusive lanes for response vehicles.

	\begin{figure}[H]
		\centering 
		\includegraphics[width=0.6\textwidth]{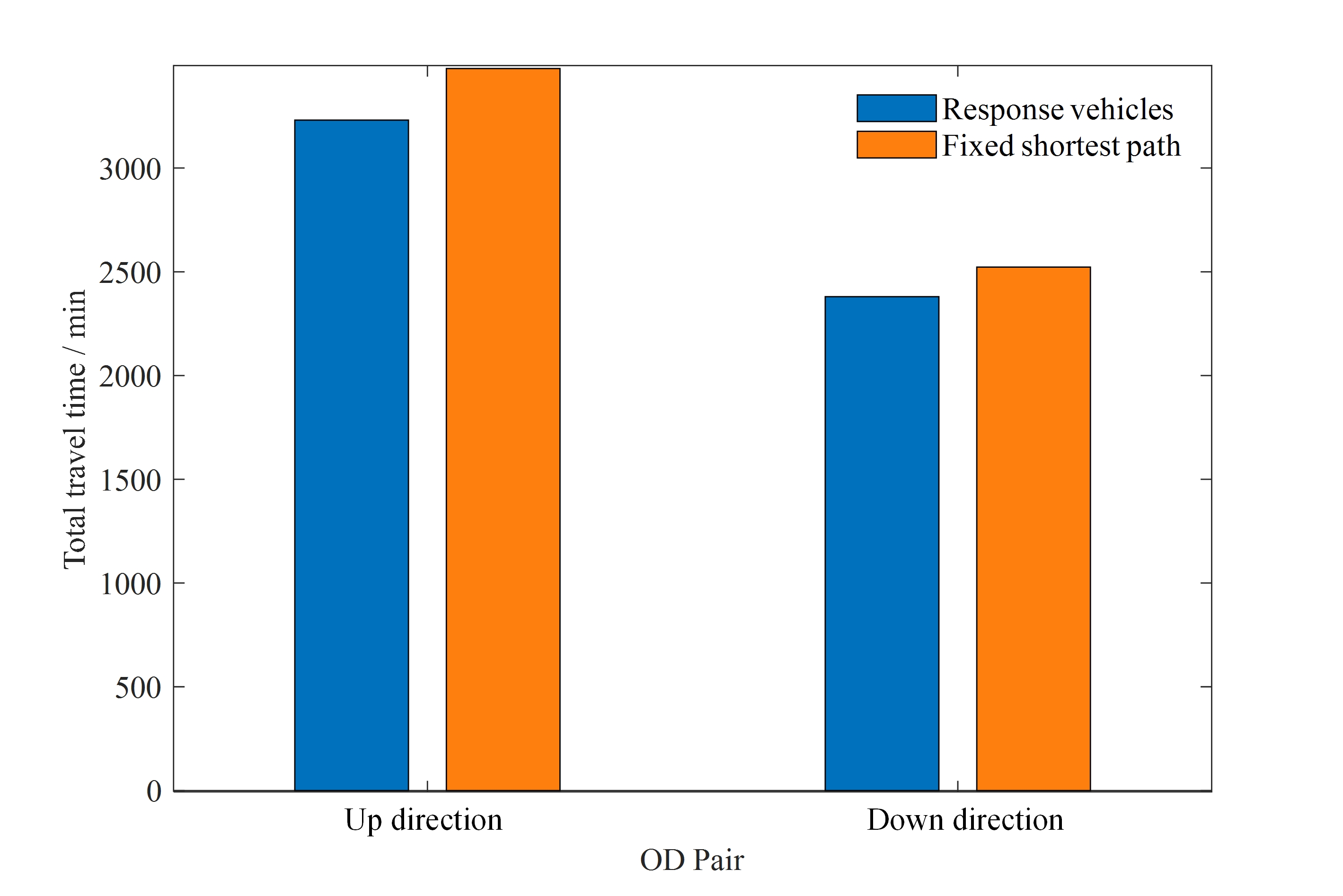}
		\caption{Comparison of total travel time of dynamic response vehicles and fixed shortest path }\label{comparison}
	\end{figure}
	\begin{figure}[H]
		\centering 
		\includegraphics[width=\textwidth]{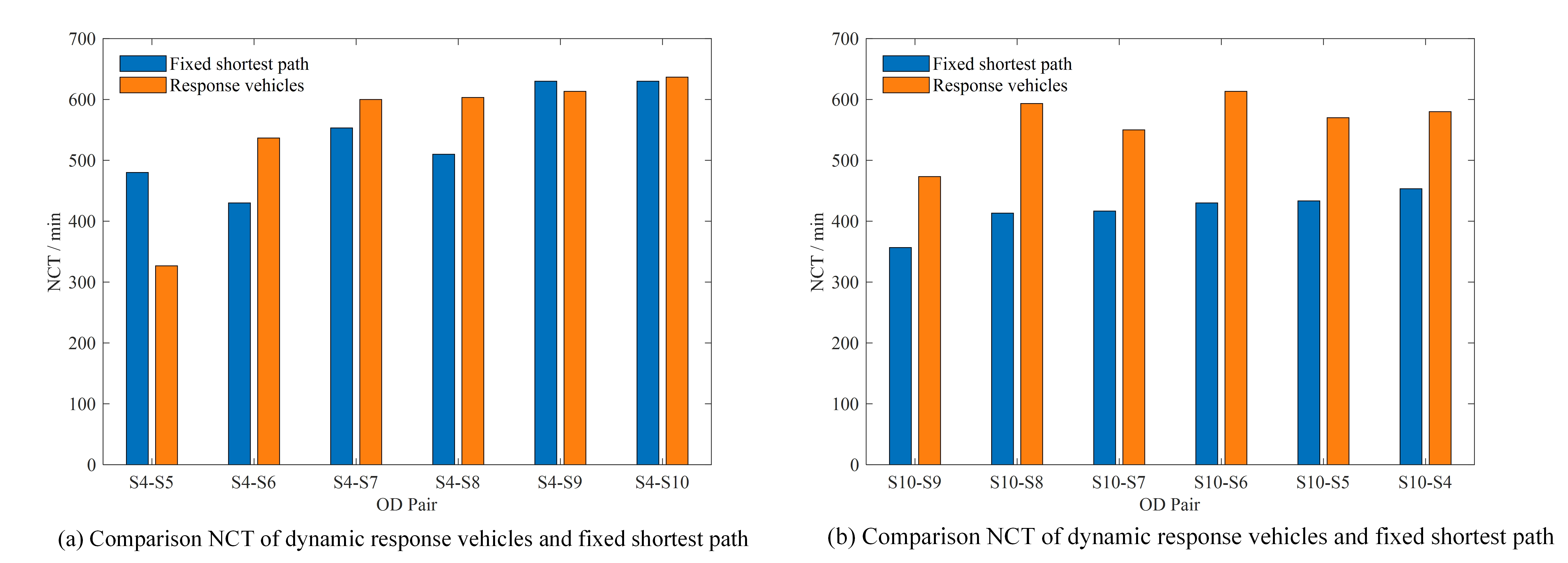}
		\caption{Comparison NCT of dynamic response vehicles and fixed shortest path }\label{comparison2}
	\end{figure}
To compare the transport efficiency of response vehicles, the total travel time of the response vehicle is compared to the shortest path scheme. The shortest path scheme sets fixed routes with the shortest distance between two stations. In this case, the shortest route is (1, 5, 9, 10, 14, 19, 20, 23, 24, 29, 30, 32, 36, 37), and vehicles are unable to change routes. The results of the response vehicles and shortest path schemes are shown in Figure \ref{comparison}, where the TTT using the response vehicles scheme is 7\% lower than the fixed shortest path scheme. However, as illustrated in Fig \ref{comparison2} (a)-(b), the NCT of different OD pairs may be greater than the fixed shortest path. Because the goal of scheduling response vehicles is to reduce total travel time for all vehicles and the NCT is determined by the capacity entering the destination station, some vehicles may stay at their original stations to avoid congestion and increase system costs.
	\begin{figure}[H]
		\centering 
		\includegraphics[width=0.8\textwidth]{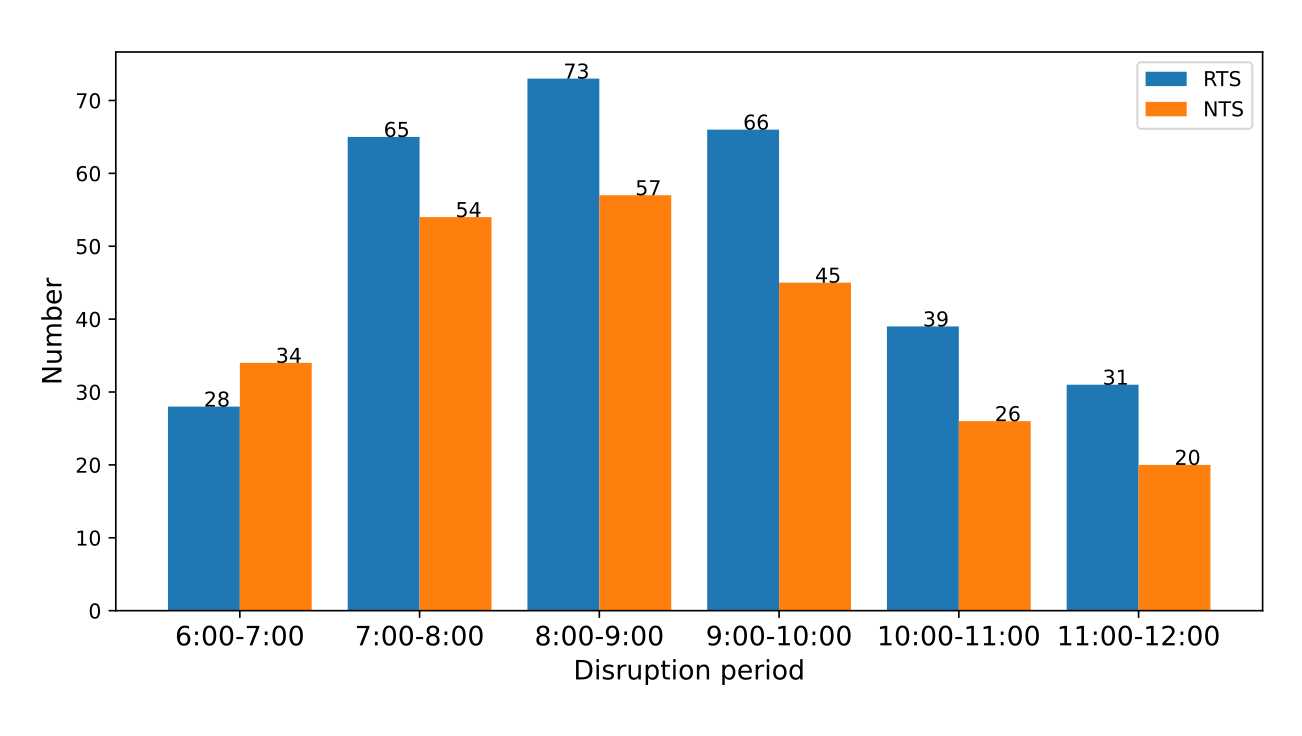}
		\caption{Number of rescheduled train services sensitive to disruption period}\label{sensitive_time}
	\end{figure}
\subsection{Sensitive analysis}
\subsubsection{timetable rescheduling sensitive to the disruption}
To investigate the phenomenon throughout different disruption periods, the disruption area is set from the FTSR station to the MM station, with a one-hour duration. Figure \ref{sensitive_time} depicts the sensitive analysis of the disruption period on the normal timetable rescheduling. Apart from the disruption period of 6:00–7:00, it can be seen that the number of rescheduled train services is greater than the number of normal train services. Early train services would be less affected by the disruption since they would have a higher chance of avoiding the spatio-temporal disruption area.  Furthermore, we can see that during other disruption periods, the disruption may not only affect normal train services during this disruption period, but also train services later in the period. This observation reflects the spread of disruptions to normal train services.
		\begin{figure}[H]
		\centering 
		\includegraphics[width=\textwidth]{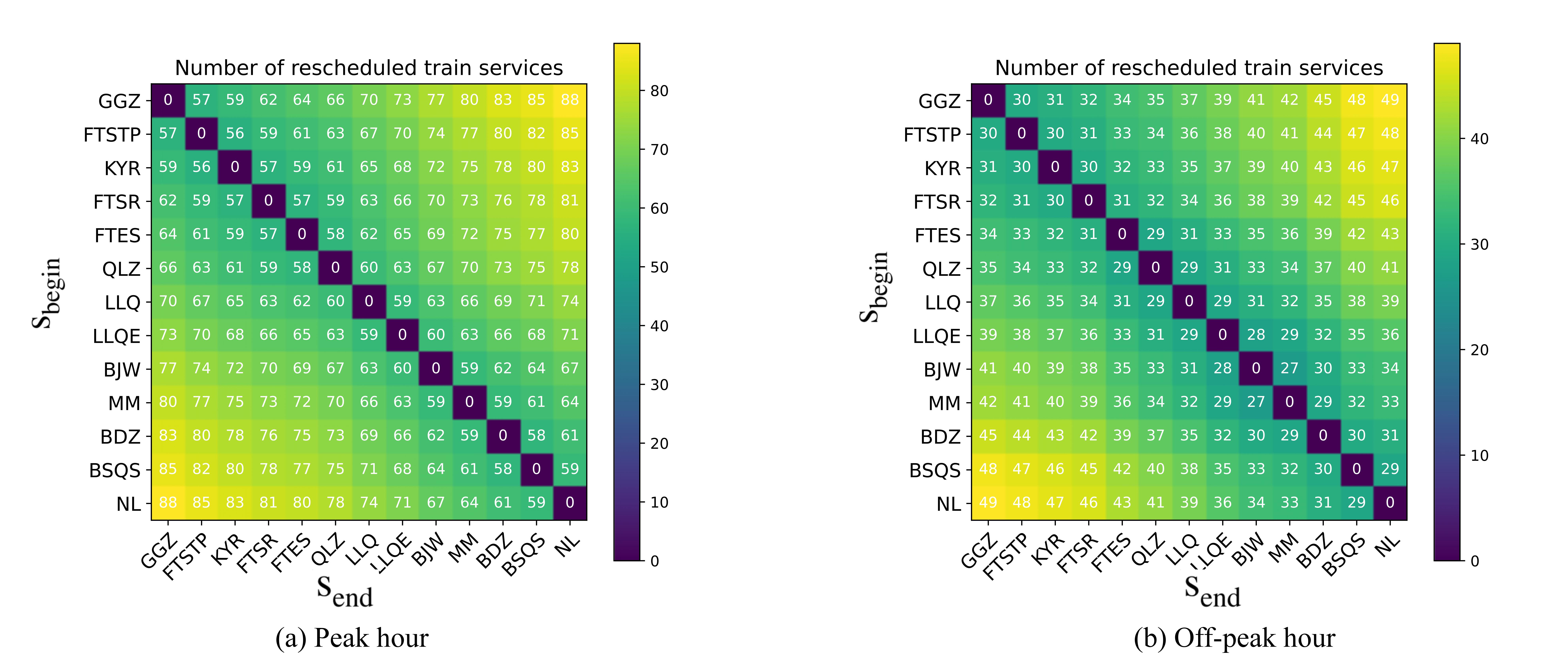}
		\caption{Number of rescheduled train services sensitive to disruption area}\label{sensitive_area2}
	\end{figure}

Furthermore, we have chosen two typical disruption periods as our focused disruption periods: peak hour (8:00–9:00 a.m.) and off-peak hour (10:00–11:00 a.m.). The sensitive analysis of the disruption period on the normal timetable rescheduling has been depicted in Figure \ref{sensitive_area2}. The number of rescheduled train services grows as the number of disrupted stations increases. Moreover, during the peak hour disruption period (see Figure \ref{sensitive_area2} (a)), more train services are rescheduled in the metro line's central area. For example, if the number of stations in the disruption area is fixed at three, the number of rescheduled train services in the disruption area from QLZ to LLQ has the greatest value, and this value decreases when the disruption area is closed to the metro line's terminal stations. During the off-peak hour disruption period (see Figure refsensitive area2 (b), more train services are rescheduled in terminal stations. 

\subsubsection{passenger accumulation sensitive to the disruption}
\begin{figure}
		\centering 
		\includegraphics[width=\textwidth]{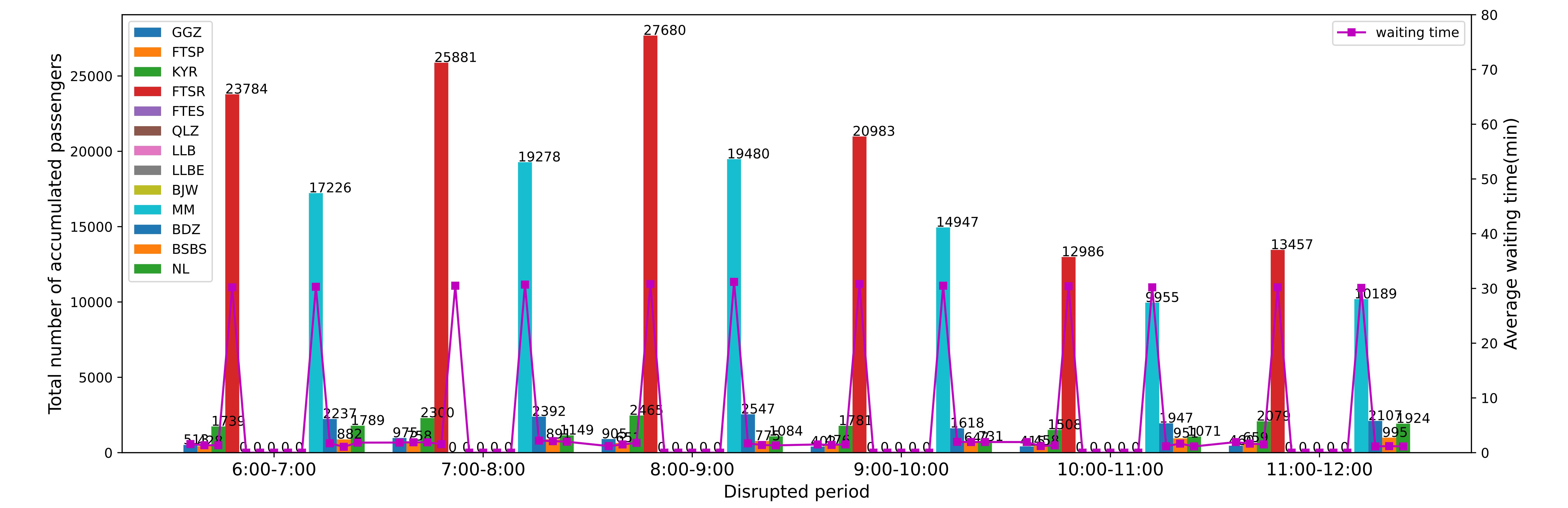}
		\caption{Passenger flow sensitive to disruption period}\label{sensitive_passenger2}
	\end{figure}

Apart from the normal metro timetable, the sensitive analysis of the disruption on passenger accumulation is also crucial in metro disruption management. Due to the time-varying passenger demands at different stations, we have set the disruption area from FTSR station to MM station, and the disruption period is one hour. As a result, the sensitive analysis of the disruption period on passenger accumulation is depicted in Figure \ref{sensitive_passenger2}.

Figure \ref{sensitive_passenger2} shows that the total number of accumulated passengers is higher before 9:00. This is due to the fact that passenger demand is usually high during the normal timetable, causing the passenger accumulation to be significantly higher during the disruption period. When an unexpected disruption occurs before 9:00 a.m., the metro manager should pay more attention. Furthermore, the average waiting time at both the FTSR station and MM station is around 30 minutes, whereas at other stations it is less than 3 minutes. This is due to the fact that passengers stranded at these two terminals will not be able to board any transportation means until the disruption is over. As a result, the response vehicles must arrive in time, or passengers may be left waiting for a long period of time on the platforms of the FTSR and MM stations.
\section{Conclusion}
Metro disruption management is currently a popular topic in metro research. In this study, we developed a two-stage optimization structure to optimize both the train timetable and response vehicle. To begin, the train schedule was adjusted to prevent conflict with the disruption area. Second, the accumulated passengers was converted into a demand for response vehicles. We used dynamic traffic assignment to simulate and optimize the scheduling of response vehicles. Moreover, if no additional train rescheduling strategies are used, this study can be viewed as a benchmark. 

This paper can be expanded in several ways. First, this paper only reschedules the train services to avoid their conflicts  with disruptions. Other train service rescheduling strategies are also possible in this paper (like more flexible short-turning, skip-stopping, etc.). Second, rolling stock circulation is also important in practical operations. Third, apart from the traffic signals discussed in our paper, more complex road conditions should be considered.
\section*{Acknowledgements}
This research was supported by the National Natural Science Foundation of China (Nos. 72288101, 71971015, 72071014) and the Fundamental Research Funds for the Central Universities (No. 2021PT206). The authors thank Su Guanghui and Sun Guofeng for providing valuable advice.
%\section*{References}
\bibliographystyle{elsarticle-num}
\bibliography{mybibfile}
\section*{Appendix}
\appendix
\section{Notations}
\setcounter{table}{0} 
\renewcommand\thetable{A.\arabic{table}}  
{\setlength{\tabcolsep}{1.5pt}
	\small
	\begin{longtable} {l}
		\caption{Notations} 
		\label{tab:test} \\
				\hline
				\textbf{Sets and indexes:}\\
				\hline
				$\mathbb{R}$ \quad Metro station set, indexed by symbol r,i.e., $\mathbb{R}=\{1,2,...,\rm{r},...,|\mathbb{R}|\}$;\\
				$\mathbb{R}_{{\rm{O}}}$ \ \  Metro station set in the operational area, indexed by symbol r;\\
				$\mathbb{R}_{{\rm{O}}}^{1}$ \ \   Metro station set in the MS-1, indexed by symbol r;\\
				$\mathbb{R}_{{\rm{O}}}^{2}$ \ \  Operational metro station in the MS-2, indexed by symbol r;\\
				$\mathbb{R}_{{\rm{D}}}$ \ \  Metro station set in disruption area, indexed by symbol r;\\
				$\mathbb{R}_{{\rm{T}}}^{1}$ \ \   First terminal station set of the disruption area in the positive direction, indexed by symbol r;\\
				$\mathbb{R}_{{\rm{T}}}^{2}$ \ \   First terminal station set of the disruption area in the negative direction, indexed by symbol r;\\
				$\mathbb{P}$ \quad Passenger group set, indexed by symbol p,i.e., $\mathbb{P}=\{1,2,...,\rm{p},...,|\mathbb{P}|\}$;\\
				$\mathbb{U}$ \ \ \ Train service in the normal timetable, indexed by symbol u,v,i.e., $\mathbb{U}=\{1,2,...,\rm{u},...,\rm{v},...|\mathbb{U}|\}$;\\
				$\mathbb{U}_{{\rm{D}}}$ \ \  Disrupted train service set during disruption period, indexed by symbol u,v,i.e., $\mathbb{U}=\{1,2,...,\rm{u},...,\rm{v},...|\mathbb{U}_{{\rm{D}}}|\}$;\\
				$\mathbb{U}_{{\rm{A}}}$ \ \ Train service set after disruption period, indexed by symbol u,v;\\
				$\mathbb{U}_{{\rm{B}}}$ \ \ Train service set before disruption period, indexed by symbol u,v;\\
				$\mathbb{T}$ \ \ \ \ Discrete time set, indexed by symbol t,i.e., $\mathbb{T}=\{1,...,\rm{t},...,|\mathbb{T}|\}$;\\
				$\mathbb{T}_{\rm{D}}$ \ \ \  Discrete time set of disruption period, indexed by symbol t;\\
				$\mathbb{G}$ \quad The cell-based evacuation network, $\mathbb{G}=(\mathbb{C},\mathbb{H})$;\\
				$\mathbb{C}$ \quad Set of cells, indexed by symbol i,i.e.,$\mathbb{C}=\{1,...,\rm{i},...,|\mathbb{C}|\}$;\\
				$\mathbb{C}_{{\rm{R}}}$ \ \ \ Set of source cells;\\
				$\mathbb{C}_{{\rm{S}}}$ \ \ \ Set of sink cells;\\
				$\mathbb{H}$\quad \ Set of intersections, indexed by symbol h,i.e., $\mathbb{H}=\{1,...,{\rm{h}},...,|\mathbb{H}|\}$;\\
				$\Gamma({\rm{i}})$ \ Set of receiving vehicles from upstream cell, indexed by symbol k;\\
				$\Gamma^{-}({\rm{i}})$ \ Set of sending vehicles to downstream cell, indexed by symbol j;\\
				$\mathbb{M}$\quad \  Set of different flow directions within each cell, indexed by m, i.e., $\mathbb{M}=\{1,2\}$;\\
				$\mathbb{T}_{\rm{R}}$ \ \ \  Discrete time set of disruption period of response vehicle, indexed by symbol t,i.e., $\mathbb{T}_{\rm{R}}=\{0,...,{\rm{t}},...,|\mathbb{T}_{\rm{R}}|\}$;\\
				$\mathbb{\tilde{T}}_{{\rm{R}}}$(t) \  The set of discrete time of time period t, indexed by symbol $\tau$.i.e., $\mathbb{\tilde{T}}_{{\rm{R}}}({\rm{t}})=\{...,\tau,...\}$.\\
				\hline
				\textbf{Parameters:}\\
				(1) parameters related to passenger flow p\\
				\hline
				$\rm{\hat{o}_{p}/\hat{d}_{p}}$  \ \ Integer parameter, the origin/destination station of passenger group p;\\
				$\rm{\hat{n}_{p}}$  \quad \ \ \ Integer parameter, the number of passenger group p;\\
				$\rm{\hat{t}_{p}}$  \quad  \ \ \ \ Real parameter, the arrival time of passenger group p at its origin station $\rm{\hat{o}_{p}}$;\\
				$\rm{\hat{f}_{p}}$   \quad   \ \ \ \ Binary parameter, wether the travel direction of passenger group p is positive, $\rm{\hat{f}_{p}}=1$ if it is, and 0 otherwise;\\
				$\rm{\tilde{o}_{p}/\tilde{d}_{p}}$  \ \ Integer parameter, the transferred origin/destination station of passenger group p;\\
				$\varphi_{{\rm{p}}}^{{\rm{u,r}}}$ \quad   \ \  Binary parameter, wether passenger flow p occupies seats at station r by taking train service u, wether $\varphi_{{\rm{p}}}^{{\rm{u,r}}}=1$ if it is,\\
				\quad \quad \quad and 0 otherwise;\\
				$\breve{{\rm{w}}}_{{\rm{p}}}^{{\rm{u}}}$ \quad   \ \   Integer parameter, the waiting time if passenger flow p takes the train service u; \\
				${\rm{\hat{w}_{u,p}}}$ \quad  Binary parameter, wether  train service u departures from $\rm{\hat{o}_{p}}$, ${\rm{\hat{w}_{u,p}}}=1$ if it is, and 0 otherwise;\\
				${\rm{\tilde{w}_{u,p}}}$ \quad Binary parameter, wether arrival time of train service u at $\rm{\hat{o}_{p}}$ is greater than $\rm{\hat{t}_{p}}$, ${\rm{\tilde{w}_{u,p}}}=1$ if it is, and 0 otherwise;\\
				${\tilde{\rm{f}}_{\rm{u,p}}}$ \quad \  Binary parameter, wether the direction of train service u and passenger flow p are same. ${\tilde{\rm{f}}_{\rm{u,p}}}=1$ if they are, and 0 otherwise;\\
				$\tilde{\theta}_{{\rm{p}}}^{{\rm{t,r}}}$ \quad \ Binary parameter, wether passenger flow p originates from station r at time t.i.e., $\tilde{\theta}_{{\rm{p}}}^{{\rm{t,r}}}=1$ if it is, and 0 otherwise;\\
				$\tilde{\vartheta}_{{\rm{p}}}^{{\rm{t,r}}}$ \quad \ Binary parameter, wether passenger flow p has arrived at station r at time t.i.e., $\tilde{\vartheta}_{{\rm{p}}}^{{\rm{t,r}}}=1$ if it does, and 0 otherwise; \\
				$\check{\theta}_{{\rm{p,u}}}^{{\rm{t,r}}}$ \quad Binary parameter, wether passenger flow p takes train service u, and demands transferring at station r at time t.i.e., $\check{\theta}_{{\rm{p,u}}}^{{\rm{t,r}}}=1$ 
				\\ \quad \quad \quad  if it is, and 0 otherwise;\\
				$\check{\vartheta}_{{\rm{p,u}}}^{{\rm{t,r}}}$ \quad Binary parameter, wether passenger flow p of taking train service u has arrived at station r at time t.i.e., $\check{\vartheta}_{{\rm{p,u}}}^{{\rm{t,r}}}=1$ if it is,
				\\ \quad \quad \quad  and 0 otherwise;\\
				$\hat{\vartheta}_{{\rm{u,p}}}^{{\rm{r,t}}}$ \quad Binary parameter, wether passenger flow p of taking train service u has departed from station r at time t.i.e., $\hat{\vartheta}_{{\rm{u,p}}}^{{\rm{r,t}}}=1$ if it is,
				\\ \quad \quad \quad and 0 otherwise;\\
				$ \xi_{\rm{p}}^{1}/ \xi_{\rm{p}}^{2}$ \ Binary parameter, wether the origin/destination of passenger flow p should be transferred;\\
				\hline
				(2) parameters related to train service\\
				\hline
				${\rm{f_{u}}}$ \quad \ \ \ \ Binary parameter, wether the travel direction of train service u is positive, ${\rm{f_{u}}}=1$ if it is, and 0 otherwise;\\
				${\rm{t}}_{{\rm{u,r}}}^{a}$/${\rm{t}}_{{\rm{u,r}}}^{d}$  \  Real parameter, the arrival/departure time of train service u at station r in original normal timetable;\\
				$\Theta_{{\rm{u}}}$ \quad \ \ Binary parameter, wether train service u conflicts with spatio-temporal disruption area, $\Theta_{{\rm{u}}}=1$ if it is, and 0 otherwise;\\
				$\delta_{{\rm{u,v}}}^{{\rm{r}}}$ \quad Binary parameter, wether train service v is the turnaround train service of u at station r. $\delta_{{\rm{u,v}}}^{{\rm{r}}}=1$ if it is, and 0 otherwise; \\
				$\theta_{{\rm{u,v}}}^{{\rm{r}}}$ \quad Binary parameter, wether train service u and v is conflicted at station r. $\theta_{{\rm{u,v}}}^{{\rm{r}}}=1$ if it is, and 0 otherwise; \\
				${\rm{\bar{C}_{u}}}$ \quad \ \ Integer parameter, the capacity of train service u;\\
				\hline
				(2) parameters related to response vehicle\\
				\hline
				${\rm{C_{R}}}$ \quad Maximum passenger holding capacity of response vehicles;\\
				${\rm{N_{i,m}^{t}}}$ \ \ Maximum number of response vehicles that can be accommodated in cell i at time t;\\
				$\alpha_{\rm{t}}$ \quad \ Length of time step t during disruption period;\\
				$\tilde{\delta}_{{\rm{i,r}}}^{{\rm{m,p}}}$ \ \ Binary parameter, wether the passenger flow p at station r matches with the transportation task (OD pair) of class-m response
				\\ \quad \quad \ vehicles departing from cell i, $\tilde{\delta}_{{\rm{i,r}}}^{{\rm{m,p}}}=1$ if it is, and 0 otherwise;\\
				${\rm{d}}_{{\rm{i,m}}}^{{\rm{t}}}$ \ \ Integer parameter, the number of m-class response vehicles originating from cell i at time t;\\
				${\rm{Q_{i,m}}}$ \ \ Integer parameter, maximum number of response vehicles that can flow into or out of cell i at time t;\\
				${{\rm{v_{m}}}}$ \quad Real parameter, free-flow velocity of m-class response vehicles;\\
				${{\rm{w_{m}}}}$ \quad Real parameter, backward shock wave propagation speed of response vehicles;\\
				\hline
				(4) other parameters\\
				\hline
				M \quad very big number;\\
				${\rm{{s}_{begin}}}/{\rm{{s}_{end}}}$ Begin/end station of disruption area;\\
				$\tau_{{\rm{begin}}}/\tau_{{\rm{end}}}$ Begin/end time of disruption period;\\
				${\rm{r}_{1}^{*}}/{\rm{r}_{2}^{*}}$ Terminal stations of disruption area;\\
				\hline
				\textbf{Decision variables:}\\
				(1)variables related to train service u\\
				\hline
				$a_{{\rm{u}}}$  \quad \quad Binary variable, wether train service u is activated;\\
				$s_{{\rm{u,r}}}$ \quad \ \  Binary variable, wether station r is visited by train service u;\\
				$t_{{\rm{u,r}}}^{a}/t_{{\rm{u,r}}}^{d}$ \ Real variable, the arrival/departure time of train service u at station r;\\
				$C_{{\rm{u}}}^{{\rm{r}}}$ \quad \ \ \ \ Integer variable, the number of passengers in train service u before it departures from station ; \\
				\hline
				(2)variables related to passenger p\\
				\hline
				$x_{{\rm{p}}}^{{\rm{u}}}$ \quad \quad Integer variable, the number of passenger flow p taking train service u;\\
				$x_{{\rm{p}}}$ \quad \quad Integer variable, the number of passenger flow p who has not boarded any train;\\
				$\mathscr{G}_{{\rm{p}}}^{{\rm{r,t}}}$\quad\ \ Integer variable, instantaneous accumulation flow p of station r at time t;\\
				$A_{{\rm{p}}}^{{\rm{r,t}}}$ \quad \ \ Integer variable, accumulated arrival flow p of station r at time t;\\
				$D_{{\rm{p}}}^{{\rm{r,t}}}$ \quad \ \ Integer variable, accumulated departure flow p of station r at time t;\\
				$G_{{\rm{p}}}^{{\rm{r,t}}}$ \quad\ \ Integer variable, accumulated stranded flow p of station r at time t;\\
				\hline
				(3)variables related to response vehicles\\
				$y_{{\rm{i,m}}}^{{\rm{t}}} $  \quad \ \ Real variable, the number of response vehicles on cell i at time t;\\
				$z_{{\rm{i,j,m}}}^{{\rm{t}}}$  \quad \ \ Real variable, the number of inflow response vehicles from cell i to cell  i+1 at a time step [t,t+1);\\
				\hline
	\end{longtable}}
\renewcommand\thefigure{B.\arabic{figure}}    
\section{Definitions}\label{definitions}
\textbf{Spatio-temporal disruption area} is a rectangle area in the train timetable, in which the train service can not go through. As shown in Figure \ref{spatio-temporal disruption area}, the spatio-temporal disruption area is bounded by the convex hull of four disruption nodes. i.e.,(${\rm{{s}_{begin},\tau_{begin}}}$), (${\rm{{s}_{end},\tau_{begin}}}$),(${\rm{{s}_{end},\tau_{end}}}$),(${\rm{{s}_{end},\tau_{end}}}$). ${\rm{{s}_{begin}}}$ and ${\rm{{s}_{end}}}$ are the terminal stations of disruption area, and $\tau_{{\rm{begin}}}/\tau_{{\rm{end}}}$ is the begin/end time of disruption. There are three train services u1,u2, and u3 representing the train service before/during/after the disruption. Train services u1 and u3 would not travel past the spatio-temporal disruption area, while train service u2 has to stop at station ${\rm{s}_{begin}}$ as it will run through the spatio-temporal disruption area. Therefore, train service u1 and u3 have not to be rescheduled, as it does not conflict with the spatio-temporal disruption area; however, reschedule strategies should be executed on train service u2 to run into the spatio-temporal disruption area.
\setcounter{figure}{0} 
\begin{figure}[H]
\centering
\includegraphics[width=0.7\textwidth]{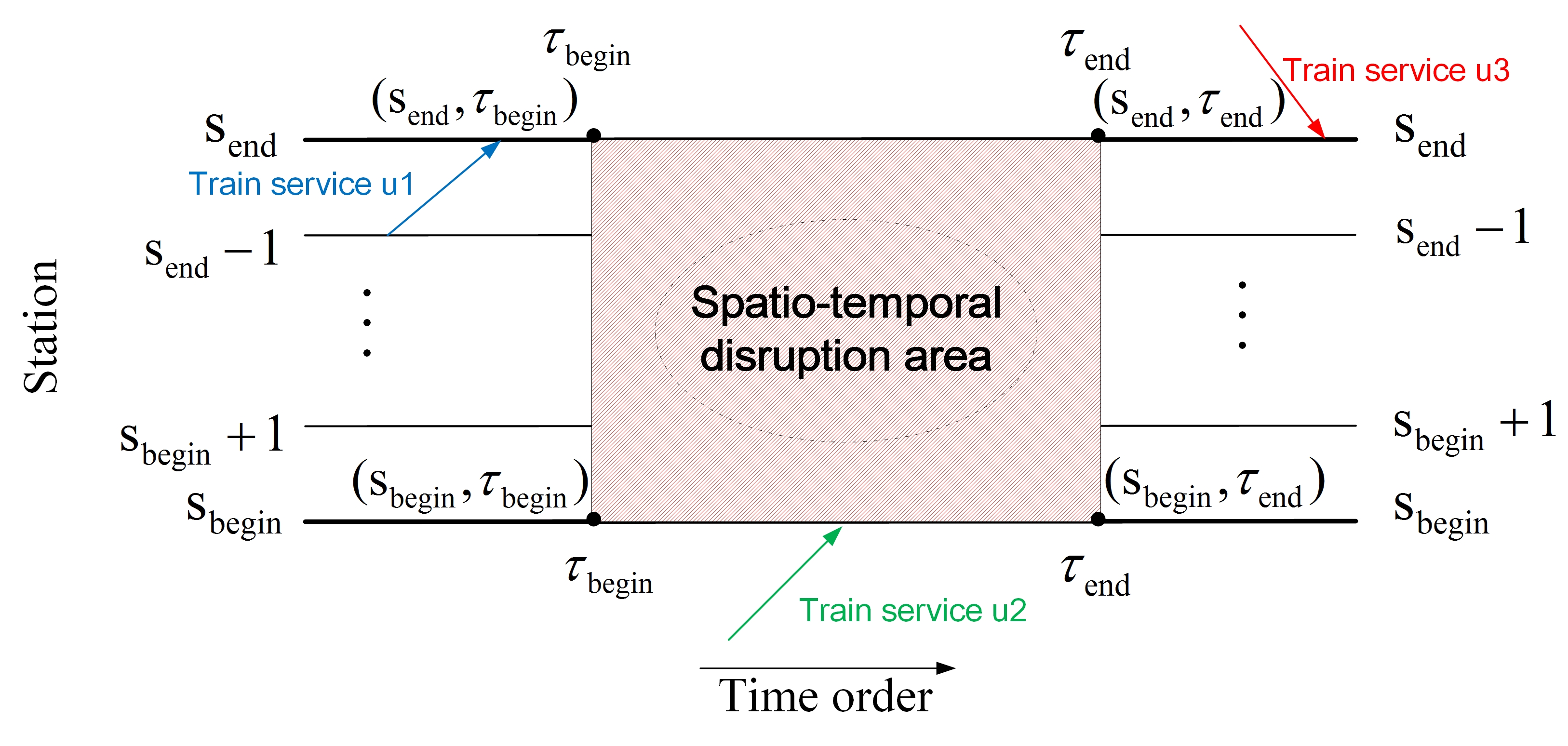}
\caption{spatio-temporal disruption area}\label{spatio-temporal disruption area}
\end{figure}
\begin{figure}[H]
\centering
\includegraphics[width=\textwidth]{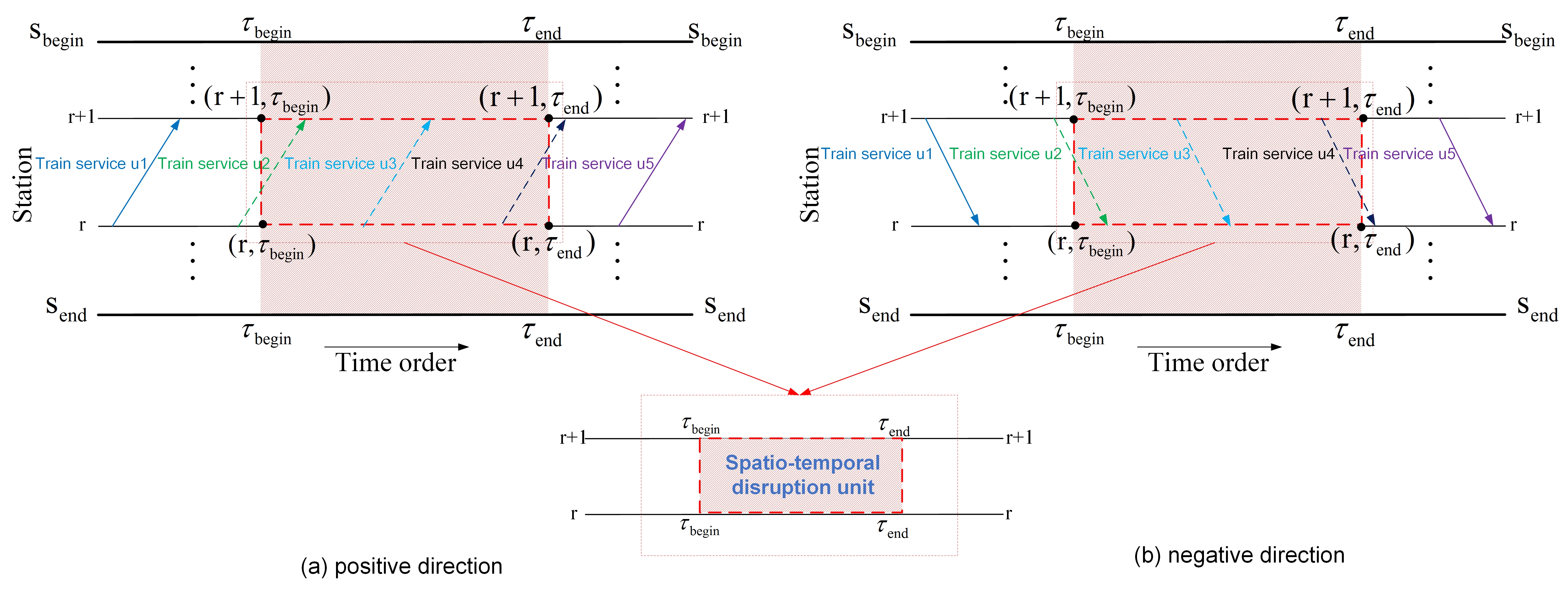}
\caption{Relation between train service and spatio-temporal disruption unit}\label{disruption_unit}
\end{figure}
\textbf{Spatio-temporal disruption unit} is a rectangle area bounded by disrupted metro line track and disruption period $\mathbb{T}_{{\rm{D}}}$. As shown in Figure \ref{disruption_unit} (a)-\ref{disruption_unit} (b), two instances have been illustrated to reveal the possible relationships between train services and spatio-temporal disruption units. Five typical train services of positive direction have traveled through line section (r, r+1), of which three train services(train service u2,u3,u4) conflict with spatio-temporal disruption unit while two train services(train service u1,u5) otherwise. By the observation of these five train services, can we see that the train service u conflicts with the spatio-temporal disruption unit if its departure/arrival time at station r/r+1 is in the range of disruption period $\mathbb{T}_{{\rm{D}}}$. Similar conclusions can also be obtained from another five train services in the negative direction (see Figure \ref{disruption_unit} (b)).Note that, the spatio-temporal disruption area is compromised of several spatio-temporal disruption units. If we want to judge whether a train service conflicts with the spatio-temporal disruption area, we have to check whether this train service conflicts with the spatio-temporal disruption units of this area one by one. Given the normal timetable and disruption information (${\rm{s_{begin}}}$,${\rm{s_{end}}}$,$\tau_{begin}$,${\rm{\tau_{end}}}$), the classification of train services can be obtained by \ref{algorithm_1}-\ref{algorithm_2}. \ref{algorithm_1} presents the classification of train services, and \ref{algorithm_2} gives the method of determining whether a train service conflicts with spatio-temporal disruption units. Additionally, the parameter $\Theta_{{\rm{u}}}$ can be determined by these two algorithms.

\section{Algorithms}
\setcounter{table}{0} 
\renewcommand\thetable{C.\arabic{table}} 
\subsection{Classification of train services}\label{algorithm_1} 
{\setlength{\tabcolsep}{1.5pt}
	\small
	\begin{longtable} {l}
		\caption{Algorithm 1: Classification of train services} 
		\label{tab:test} \\
				\hline
				Algorithm 1: Classification of train services\\
				\hline
				00: (\textbf{Initialization}) $\mathbb{U}_{{\rm{B}}}$, $\mathbb{U}_{{\rm{D}}}$, $\mathbb{U}_{{\rm{A}}}:=\Phi$\\
				01: For u $\in \mathbb{U}$\\
				02: \quad If train service u does not conflict with spatio-temporal disruption area( judge by Algorithm 2.)\\
				03: \qquad    If the arrival/departure time of train service u at its origin station $t_{{\rm{u,r}}}^{a}> \tau_{\rm{end}}||t_{{\rm{u,r}}}^{d}> \tau_{\rm{end}}$\\
				04: \qquad \quad  $\mathbb{U}_{{\rm{A}}} \gets$ u;\\
				05: \qquad    Else if the arrival/departure time of train service u at its origin station $t_{{\rm{u,r}}}^{a}< \tau_{\rm{begin}}||t_{{\rm{u,r}}}^{d}< \tau_{\rm{begin}}$\\
				06: \qquad \quad  $\mathbb{U}_{{\rm{B}}} \gets$ u;\\
				07: \qquad    Else \\
				08: \qquad \quad  $\mathbb{U}_{{\rm{O}}} \gets$ u;\\
				09: \qquad    End if \\
				10: \quad  Else\\
				11: \qquad \quad  $\mathbb{U}_{{\rm{O}}} \gets$ u;\\
				11: \qquad \quad  Adjust the arrival/departure time of train service u at each station r(r $\in \mathbb{R}$) $t_{{\rm{u,r}}}^{a}/t_{{\rm{u,r}}}^{d}$ by Algorithm 2;\\
				12: \quad End if\\
				13: End for \\
				14: \textbf{Output:}$\mathbb{U}_{{\rm{B}}}$,$\mathbb{U}_{{\rm{O}}}$, $\mathbb{U}_{{\rm{A}}}$;\\
				\hline
	\end{longtable}}
\subsection{Detection of the conflict between a train service and spatio-temporal area} \label{algorithm_2} 
{\setlength{\tabcolsep}{1.5pt}
	\small
	\begin{longtable} {l}
		\caption{Algorithm 2: Detection of the conflict between a train service and spatio-temporal area} 
		\label{tab:test} \\
				\hline
				Algorithm 2: Detection of the conflict between a train service and spatio-temporal area\\
				\hline
				00: \textbf{Input:} train service u and its station arrival/departure time ${\rm{t}}_{{\rm{u,r}}}^{a}/{\rm{t}}_{{\rm{u,r}}}^{d}$ in the original normal timetable \\
				01:(\textbf{Initialization}) $\Theta_{{\rm{u}}}:=0$;\\
				02:  For r $\in [{\rm{{s}_{begin}}}, {\rm{{s}_{end}}}]$\\
				03:  \quad If  ${\rm{t}}_{{\rm{u,r}}}^{a}\in ( \tau_{\rm{begin}}, \tau_{\rm{end}})||{\rm{t}}_{{\rm{u,r}}}^{d} \in (\tau_{\rm{begin}}, \tau_{\rm{end}})$;\\
				04: \quad \quad $\Theta_{{\rm{u}}}=1$;\\
				05:  \quad End if\\
				06:   End for \\
				07: \textbf{Output:}$\Theta_{{\rm{u}}}$;\\
				\hline
	\end{longtable}}
	Note that, the detection of the conflict between a train service u and spatio-temporal area is judged by the relation between the arrival/departure time of train service u at each station of the disruption area and the disruption period $[{\rm{{s}_{begin}}}, {\rm{{s}_{end}}}]$. If there is a time belonging to the disruption period, then train service u conflicts with the spatio-temporal area ($\Theta_{{\rm{u}}}=1$), and $\Theta_{{\rm{u}}}=0$ otherwise.
\subsection{Dynamic filling algorithm} \label{algorithm_3} 
	{\setlength{\tabcolsep}{1.5pt}
	\small
	\begin{longtable} {l}
		\caption{Algorithm 3: Dynamic filling algorithm} 
		\label{tab:test} \\
				\hline
				Algorithm 3: Dynamic filling algorithm\\
				\hline
				00: (\textbf{Initialization}) Let $\varphi_{{\rm{p}}}^{{\rm{u,r}}}$ be a $|\mathbb{P}|\times|\mathbb{U}|\times|\mathbb{R}|$ zero matrix, and input related parameters.\\
				01: For u  $\in \mathbb{U}$\\
				02: \quad If $\Theta_{{\rm{u}}}=1$\\
				03: \quad \quad For p $\in \mathbb{P}$\\
				04: \quad\quad\quad If ${\rm{\hat{f}_{p}}}=1$\\
				05:  \quad  \quad\quad\quad For r=${\rm{\hat{o}_{p}}}:{\rm{\hat{d}_{p}}}-1$\\
				06:  \quad  \quad  \quad\quad\quad  If r $\ne{\rm{{s}_{begin}}}$ \\
				07:  \quad  \quad  \quad \quad\quad\quad $\varphi_{{\rm{p}}}^{{\rm{u,r}}}=1$;\\
				08: \quad  \quad  \quad\quad\quad  Else\\
				09:  \quad  \quad  \quad\quad \quad\quad $\varphi_{{\rm{p}}}^{{\rm{u,r}}}=1$, break;\\
				10:  \quad  \quad  \quad\quad\quad  End if\\
				11:  \quad  \quad\quad\quad End for \\
				12: \quad \quad\quad Else \\
				13:  \quad  \quad\quad\quad For r=${\rm{\hat{d}_{p}}}+1:{\rm{\hat{o}_{p}}}$\\
				14:  \quad  \quad  \quad\quad\quad  If r $\ne{\rm{{s}_{end}}}$ \\
				15:  \quad  \quad  \quad \quad\quad\quad $\varphi_{{\rm{p}}}^{{\rm{u,r}}}=1$;\\
				16: \quad  \quad  \quad\quad\quad Else\\
				17:  \quad  \quad  \quad\quad \quad\quad $\varphi_{{\rm{p}}}^{{\rm{u,r}}}=1$, break;\\
				18:  \quad  \quad  \quad\quad\quad  End if\\
				19:  \quad  \quad\quad\quad End for \\
				20:  \quad  \quad\quad End if \\
				21:  \quad  \quad End for \\
				22: \quad Else\\
				23: \quad \quad For p $\in \mathbb{P}$\\
				24: \quad\quad\quad If ${\rm{\hat{f}_{p}}}=1$\\
				25:  \quad  \quad\quad\quad For r=${\rm{\hat{o}_{p}}}:{\rm{\hat{d}_{p}}}-1$\\
				26:  \quad  \quad  \quad \quad\quad $\varphi_{{\rm{p}}}^{{\rm{u,r}}}=1$;\\
				27:  \quad  \quad\quad\quad End For\\
				28:  \quad  \quad\quad Else\\
				29:  \quad  \quad\quad\quad For r=${\rm{\hat{d}_{p}}}+1: {\rm{\hat{o}_{p}}}$\\
				30:  \quad  \quad  \quad \quad\quad $\varphi_{{\rm{p}}}^{{\rm{u,r}}}=1$;\\
				31:  \quad  \quad\quad\quad End for\\
				32: \quad\quad\quad End If\\
				33: \quad\quad End for\\
				34: \quad End if\\
				35:  End for \\
				36: \textbf{Output:}$\varphi_{{\rm{p}}}^{{\rm{r}}}$;\\
				\hline
	\end{longtable}}
	Note that, the identification of passenger holding is distinguished by the train service. If the train service u conflicts with the spatio-temporal area, then the value of $\varphi_{{\rm{p}}}^{{\rm{u,r}}}$ would be filled by a dynamic filling procedure. that is, the parameter $\varphi_{{\rm{p}}}^{{\rm{u,r}}}$ is set to one from his origin station to the terminal station of the disruption area; however, if the train service u does not conflict with spatio-temporal area, then the value of $\varphi_{{\rm{p}}}^{{\rm{u,r}}}$ is set to one along the stations from origin station of passenger flow p to station before his destination station.
\section{Figures}\label{appendix_figure}
\setcounter{figure}{0} 
\renewcommand\thefigure{D.\arabic{figure}} 
	 	\begin{figure}[H]
		\centering 
		\includegraphics[width=0.9\textwidth]{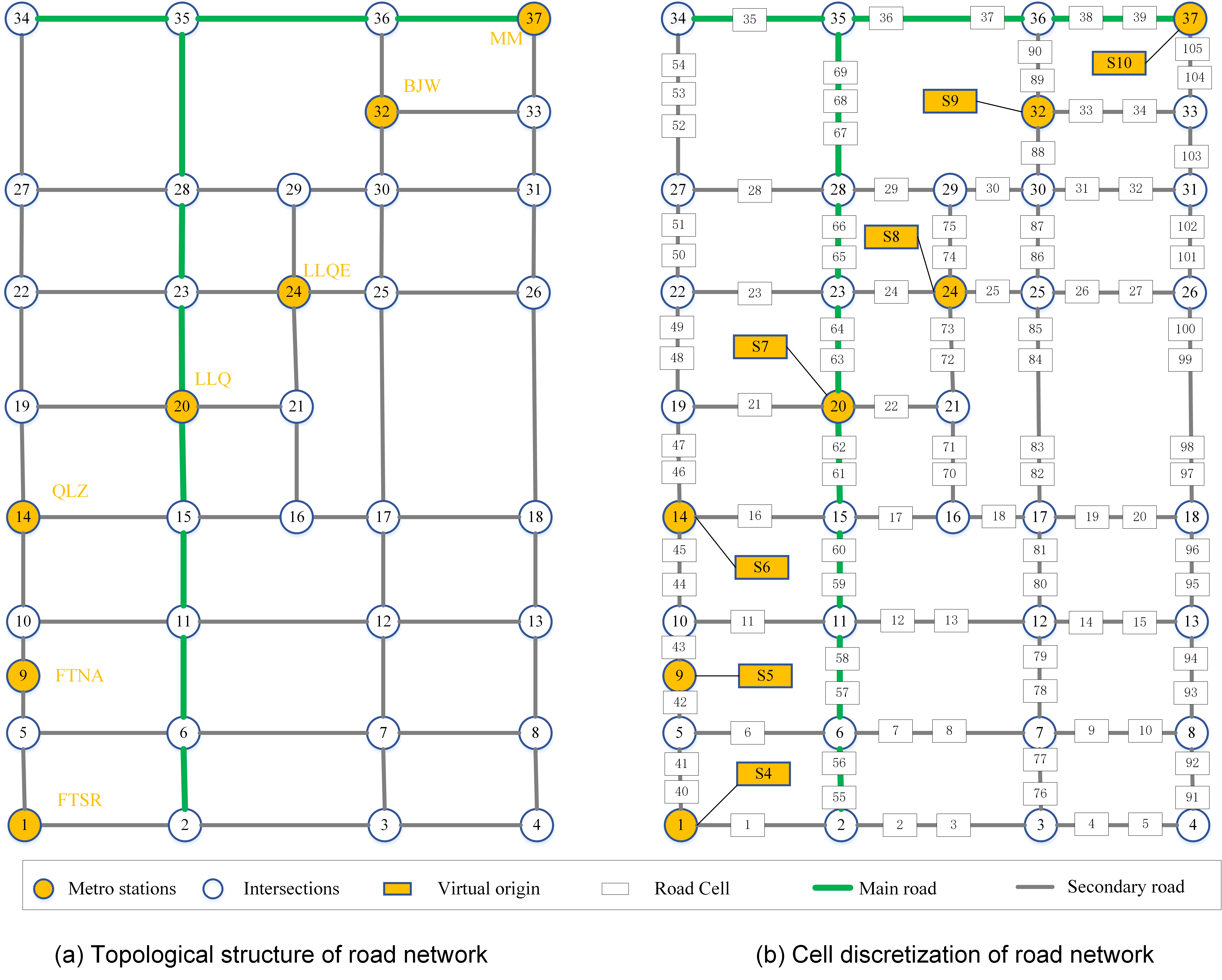}
		\caption{Road network}\label{road_network}
	\end{figure}
\section{Proofs}
\subsection{Proof of Remark 1.}\label{proof1}
We first prove that if M $\ge \max\limits_{{\rm{p}\in \mathbb{P}}, {\rm{u}}\in \mathbb{U}\cup\mathbb{\tilde{U}}}\{\breve{{\rm{w}}}_{{\rm{p}}}^{{\rm{u}}}\}$, then the passengers can be assigned completely. For any passenger flow p (${\rm{p}\in \mathbb{P}}$), if this passenger flow is assigned into a feasible train service u(${\rm{u}\in \mathbb{U}\cup\mathbb{\tilde{U}}}$), then the waiting time of this passenger flow would be $z_{1}=\breve{{\rm{w}}}_{{\rm{p}}}^{{\rm{u}}}$; Otherwise, if it does not be assigned into any trains, then the penalty cost would be $z_{2}={\rm{M}}$. It is obvious that $z_{1}\le z_{2}$ if M $\ge \max\limits_{{\rm{p}\in \mathbb{P}}, {\rm{u}}\in \mathbb{U}\cup\mathbb{\tilde{U}}}\{\breve{{\rm{w}}}_{{\rm{p}}}^{{\rm{u}}}\}$. Therefore, to minimize the objective function, the passengers are ought to be assigned to its feasible train services.

Secondly, we give an example to prove that if M<$\max\limits_{{\rm{p}\in \mathbb{P}}, {\rm{u}}\in \mathbb{U}\cup\mathbb{\tilde{U}}}\{\breve{{\rm{w}}}_{{\rm{p}}}^{{\rm{u}}}\}$, then the passengers would not be assigned completely. For any passenger flow p(${\rm{p}\in \mathbb{P}}$), let
M $\le \max\limits_{{\rm{u}}\in \mathbb{U}\cup\mathbb{\tilde{U}}}\{\breve{{\rm{w}}}_{{\rm{p}}}^{{\rm{u}}}\}$, then the passenger group p is better to be stranded, as the stranded cost(M) is less than the waiting time of any train service u($\breve{{\rm{w}}}_{{\rm{p}}}^{{\rm{u}}}$).Q.E.D.

\end{document}